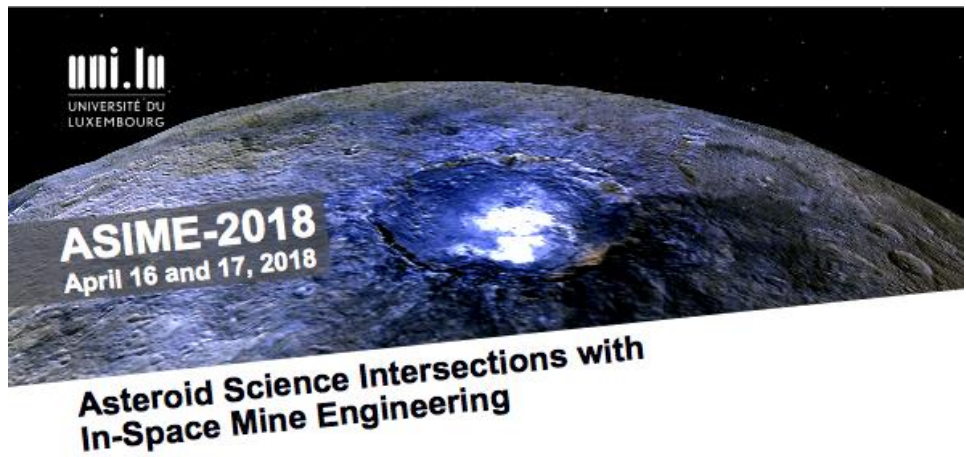

**ASIME 2018 White Paper**
**IN-SPACE UTILISATION OF ASTEROIDS:**
**Asteroid Composition**
**Answers to Questions from the Asteroid Miners**

**by Amara Graps + 43 Co-Authors**

Outcome from the ASIME 2018: Asteroid Intersections with Mine Engineering, Luxembourg. April 16-17, 2018.

**Date: 25.04.2019**
**Version 1.2**

# AUTHORS

| Amara L. Graps **lead author** | for correspondence: **graps@psi.edu** ) |
| --- | --- |
| Angel Abbud-Madrid | Colorado School of Mines, USA |
| Paul Abell | NASA |
| Antonella Barucci | Observatoire Paris-Site de Meudon |
| Pierre Beck | University of Grenoble, France |
| Lydie Bonal | University of Grenoble, France |
| Grant Bonin | Deep Space Industries, Inc., USA |
| Øystein Risan Borgersen | SolSys Mining AS |
| Daniel Britt | University of Central Florida, Orlando, FL, USA |
| Humberto Campins | University of Central Florida, Orlando, FL, USA |
| Kevin Cannon | University of Central Florida, Orlando, FL, USA |
| Ian Carnelli | ESA |
| Benoît Carry | Observatoire de la Côte d'Azur |
| Ian Crawford | Birkbeck College, University of London |
| Julia de Leon | Instituto de Astrofisica de Andalucia – CSIC, Granada, Spain |
| Line Drube | German Aerospace Center, Koeln, Germany |
| Kerri Donaldson-Hanna | University of Central Florida, Orlando, FL, USA |
| Martin Elvis | Harvard-Smithsonian Center for Astrophysics, Cambridge, MA, USA |
| Alan Fitzsimmons | Harvard-Smithsonian Center for Astrophysics, Cambridge, MA, USA |
| JL Galache | Aten Engineering |
| Simon F. Green | School of Physical Sciences, The Open University, Milton Keynes, UK |
| Jan Thimo Grundmann | German Aerospace Center, Koeln, Germany |
| Alan Herique | Univ. Grenoble Alpes, France |
| Daniel Hestroffer | Observatoire de Paris, France |
| Henry Hsieh | Planetary Sciences Institute, Tucson, AZ USA |
| Akos Kereszturi | Hungarian Academy of Sciences, Budapest, Hungary |
| Michael Kueppers | ESA |
| Chris Lewicki | Planetary Resources / ConSynsus |
| Yangting Lin | Institute of Geology and Geophysics, China |
| Amy Mainzer | JPL, USA |
| Patrick Michel | Observatoire de la Côte d'Azur, Nice, France |
| Hong-Kyu Moon | Korea Astronomy and Space Science Institute |
| Tomoki Nakamura | Tohoku University |
| Antti.i Penttila | University of Helsinki, Finland |
| Sampsa Pursiainen | Tampere University of Technology , Finland |
| Carol Raymond | JPL, USA |
| Vishnu Reddy | LPL, University of Arizona, Tucson, Arizona, USA |
| Andy Rivkin | Johns Hopkins University Applied Physics Laboratory, USA |
| Joel Sercel | Trans Astra, Los Angeles, USA |
| Angela Stickle | Johns Hopkins University Applied Physics Laboratory, USA |
| Paolo Tanga | Observatoire de la Côte d'Azur, Nice, France |
| Mika Takala | Tampere University of Technology, Finland |
| Tom Wirtz | Luxembourg Institute of Science and Technology |
| YunZhao Wu | Key Laboratory of Planetary Sciences, Purple Mountain Observatory, Chinese Academy of Sciences, Nanjing 210034, China |

The Companies provided questions: Aten Engineering, Planetary Resources, and Deep Space Industries

# Contents

Amara Graps | graps@psi.edu

Amara Graps | graps@psi.edu

# Introduction

In keeping with the Luxembourg government's initiative to support the future use of space resources, a second workshop following ASIME 2016 was held in Belval, Luxembourg on April 16-17, 2018.

The previous ASIME 2016 provided an environment for the detailed discussion of the specific properties of asteroids, with the engineering needs of space missions that utilize asteroids. It produced a layered record of discussions from the asteroid scientists and the asteroid miners to understand each other's key concerns and to address key scientific questions from the asteroid mining companies. It resulted in a White Paper (Graps et al, 2016) that points to the science knowledge gaps: "SKGs" for advancing the asteroid in-space resource utilisation domain.

The goal of ASIME 2018 was to focus on one of those SKGs: **asteroid composition**.

> *What do we know about asteroid composition from remote-sensing observations? What are the potential caveats in the interpretation of Earth-based spectral observations? What are the next steps to improve our knowledge on asteroid composition by means of ground-based and space-based observations and asteroid rendez-vous and sample return missions?  How can asteroid mining companies use this knowledge?*

The goal of ASIME 2018 two-day workshop of almost 70 scientists and engineers was to have detailed discussions on this focused topic, in the context of the engineering needs of space missions with in-space asteroid utilisation.

We were a diverse group. Out of **68 attendees**, the percentage of: **industry participants: 26%**, **student participants: 10%**, **female speakers: 20%**,  **female keynote speakers: 39%**.  The participants represented **14 countries** including the USA, China, Japan, and S. Korea.

Amara Graps | graps@psi.edu

# Current Science Knowledge Gaps / Needs

The following are derivative concepts: the **Science Knowledge Gaps (SKG)**, from the 2016 Answers to the Questions from the Asteroid Miners.

*Especially Meteorite links to C-type asteroids*

**1.** **More studies are needed to map the classification of meteorites to asteroids.** Presently the best-established link is between ordinary chondrites and S-type asteroids. We need more useful published literature about the bulk composition of meteorites to help make more accurate simulants. We need to understand the meteorite links to C-type asteroids.

*This point was raised again often.*

**2. Dedicated NEA discovery and follow-up instrumentation.** The best observability conditions for a given NEA are typically offered around the discovery time (brightest). Need to run observations to characterize NEAs quickly after discovery; best possible with dedicated telescope(s). What is needed: A photometric telescope of a 2-3m class (to reach V ~ 21 with good S/N) available on short notice (for that the observations can be best taken right after discovery). To characterize one NEA, with full IR/Vis spectral characterizations, but with 'proxies' or short cuts to 'each NEO'.

*Regolith*

**3.** **An understanding of granular material dynamics in low-gravity.** Before being sure that we have a robust understanding of the asteroid regolith and to seriously start some systematic material extraction / utilization programs, we must understand how this regolith with its properties responds to the envisage action, i.e. to understand granular material dynamics in low-gravity. Missions like Hera, Hayabusa 2 and OSIRIS-REx will address.

*Map of "low-albedo NEOs" (presumably dark and carbonaceous)*

**4.** **Identifying the available low-delta-v (which are the objects with orbits similar to the Earth) targets is key.** What is needed is a map of low delta-v, low synodic period and low-albedo NEOs as a first-cut to fine-tune the target possibilities.

*Orbital Family clues to Asteroid Composition*

**5.** **Determine if a NEO's dynamically predicted source regions is consistent with its actual physical characterizations.** Knowing the asteroid's source region, and hence, it's orbital family characteristics, can enable a short-cut to characterize the small NEOs of that family which are difficult to measure spectroscopically.

*Regolith*

**6.** **For making useful asteroid regolith simulants, immediate needs are: adequate data on the particle sizing of asteroid regolith and sub-asteroid-regolith surface.** How does the asteroid regolith vary with depth? If the NEOs have structure like comet nucleus 67P, then the NEO regolith is denser than the deep interior.

*The 21 Questions have been ordered more logically than before with suggestions from Alan Fitzsimmons and Simon Green, and for 2018, placed in an alphabetic-ordering.*

# I. Potential Targets
Spectroscopic and Photometric reconnaissance

**Number of NEOs and Size Distribution?**

a) Size distribution (Green, 2016)
Discovery and re-observation statistics + observation bias models give good indication of NEO population
D>1 km: ~1000;  D>100 m : 35 000;  D>10 m: ~ 5 million  (Tricario, 2016)

b)  Elvis (2016-ASIME White Paper) says: ~20,000 NEOs >100m (NEOWISE) and ~20 million NEOs > 20 meters (bolide, infrasound). The shape of the number-size curve in this very steep range is not known and clearly makes a big difference. Efforts are underway with the DECam in Chile to determine this.

c) (Galache, 2016)

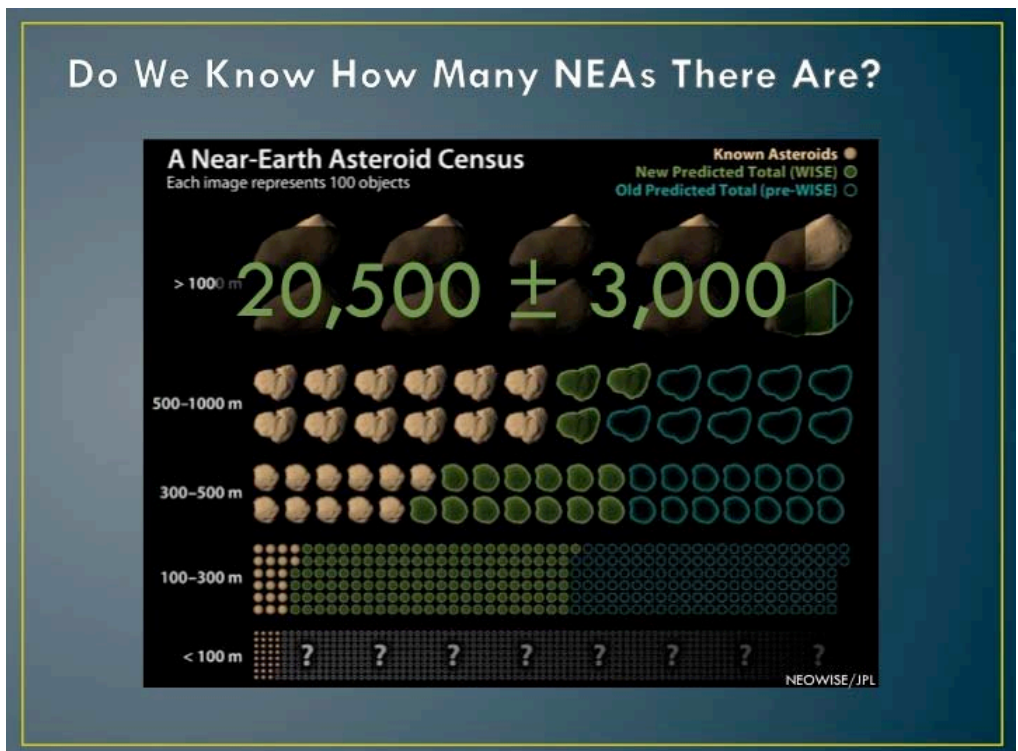

**Figure 1.** How many NEAs are there? From NEOWISE/JPL and Galache, 2016.

<u>To make the spectral assessment accurately</u>. the best way is *not* to rely on spectral data (which is hard to acquire), but instead, depend on large surveys such as GAIA and NEOWISE and use their colors/albedos as proxies for composition See (Carry, 2018).

(M. Mueller), Some information on the taxonomic type can be gleaned from measurements of the albedo (optical reflectivity). In particular, there's a one-to-one correspondence between primitive (and presumably water-rich) composition and low albedo (geometric albedo < ~7.5%). Note that, while hydrated minerals are almost always found on low-albedo asteroids, finding a low-albedo asteroid is not sufficient to guarantee hydrated minerals will be present. For instance, the meteorite Allende is anhydrous for these purposes and is a low albedo carbonaceous chondrite, and the Almahatta Sitta meteorite is low albedo and igneous.

Caveat for Gaia: It is severely limited in magnitude, as it can't go deeper than V~20 (the real threshold is around 20.5). As a consequence, it can provide spectra for a limited number of NEOs. On the other hand, Gaia's spectral resolution (equivalent to ~30 independent bands) is a good match with the asteroid characterization requirements, but it's limited to the visible range. Diagnostic bands at > 1 micron are out of reach and require complementary observations

<u>(Green, 2018). With an eye towards a composition priority of **water**</u>.

**Caveat**: *Not all* of the asteroid mining companies have this priority. G. Bonin (DSI) said in the Green Wrap-up that Water was NOT the initial goal for DSI.

- Spectroscopy: ~2500 asteroids in visible, many fewer in IR
- Spectrophotometry: ~60 000 in 3-8 wavebands
- Wide diversity of spectral types in NEO population
- Primitive objects have relatively featureless spectra
- Presence of hydrated minerals can be identified
  - Through 3 micron feature
  - In thermal IR
  - By 0.7 micron feature (not exclusive)
- Growing understanding of affects of space weathering on low albedo asteroid spectra

Amara Graps | graps@psi.edu

# Water in Asteroids

- 70% of C-type asteroids show the presence of aqueous altered minerals

Analysis on a sample of **625** visible spectra belonging to primitive classes shows the large presence of aqueous altered minerals.

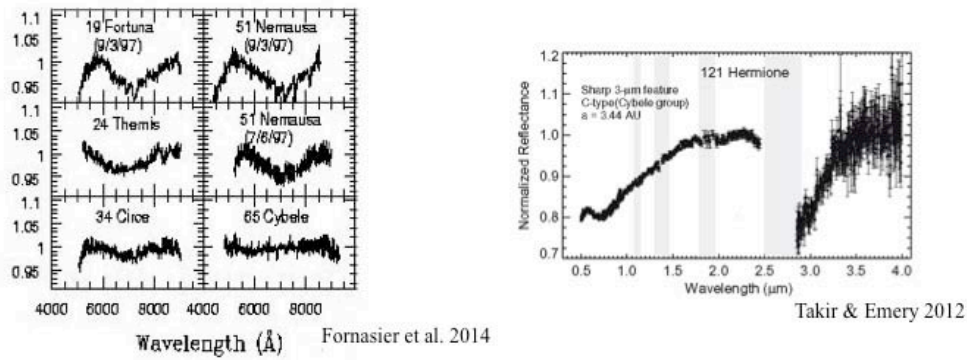

**Figure 2.** From Barucci, 2018-ASIME.

## AQUEOUS ALTERATION ON PRIMITIVE ASTEROIDS

| λ (μm) | width (μm) | Transition |
|---|---|---|
| <0.4 | >0.1 | $Fe^{2+} \rightarrow Fe^{3+}$ intervalence charge |
| 0.43 | 0.02 | $Fe^{3+}$ spin forbidden (as in jarosite) |
| 0.60-0.65 | 0.12 | $6AI \rightarrow 4T2(G)\ Fe^{3+}$ in Fe alteration minerals |
| 0.70 | 0.30 | $Fe^{2+} \rightarrow Fe^{3+}$ in phyllosilicates |
| 0.80-0.90 | 0.08 | $6AI \rightarrow 4T1(G)\ Fe^{3+}$ in Fe alteration minerals |
| 3.0 | >0.7 | Structural OH interlayer and adsorbed $H_2O$ |
| 3.07 | 0.2 | $H_2O$ ice, $NH_4$ bearing saponite |

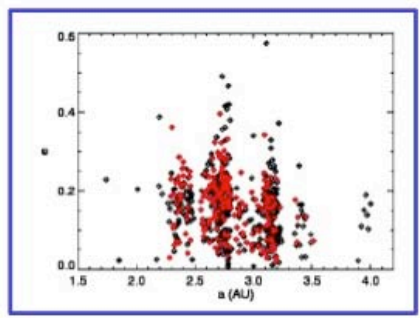

Vilas et al., 1993, 1994
Barucci et al., 1998

**~70% of C-type**

Fornasier et al., 2014

**Figure 3.** From Barucci, 2018-ASIME.

Rivkin (2018-ASIME) provided more statistics of water-rich NEOs:

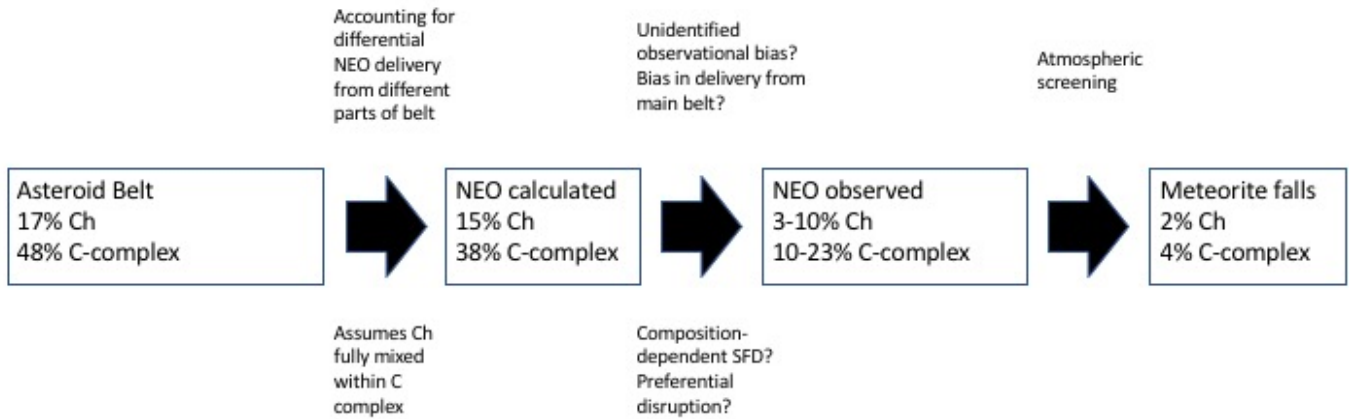

**Figure 4.** From Rivkin, 2018-ASIME.

| Population | Total Known Number | Estimated number of Ch asteroids in population |
|---|---|---|
| NEOs > 1 km | 886 | ~25-100 |
| NEOs > 1 km, Dv < 7.2 km/s | 152 | ~5-18 |
| NEOs > 100 m, Dv < 7.2 km/s | 2764 | ~80-330 |
| All NEOs with Dv < 7.2 km/s | >9500 | ~280-1000+ |
| NEOs > 100 m, Dv < 5.1 km/s | ~1100 | ~5-6 |

- NEO delivery models estimate a factor of 10 more Ch asteroids than meteorite falls would suggest.
- Estimates derived from NEO observations suggest 3-12% should be Ch class
- Statistically, at least 75 currently-known asteroids > 100 m diameter should* be Ch class and more accessible than the lunar surface.
- Ongoing surveys may identify these objects, but a dedicated observing program to identify these asteroids as Ch might be necessary*

For more discoveries: There exists in the archives already a lot of data. (Green, 2018; Carry, 2018). Needs careful selection and follow-up.

- Dedicated surveys
  - >18 000 NEAs discovered; >1000 potential targets for us
  - >50% smaller than 140 m; sizes down to a few metres
  - We can identify accessible targets: low delta-v (low *e, i*; *a* ~1 AU)

- New surveys (Gaia, LSST) will increase discoveries
  - May be serendipitous data in other surveys

Amara Graps | graps@psi.edu

- >50% smaller than 140 m; sizes down to a few metres

See 2018-ASIME Wrap-Up talk April 17, 2018: Green, 2018
(https://youtu.be/9SX4nFnIN4M?t=25173 )

For example, the SDSS.

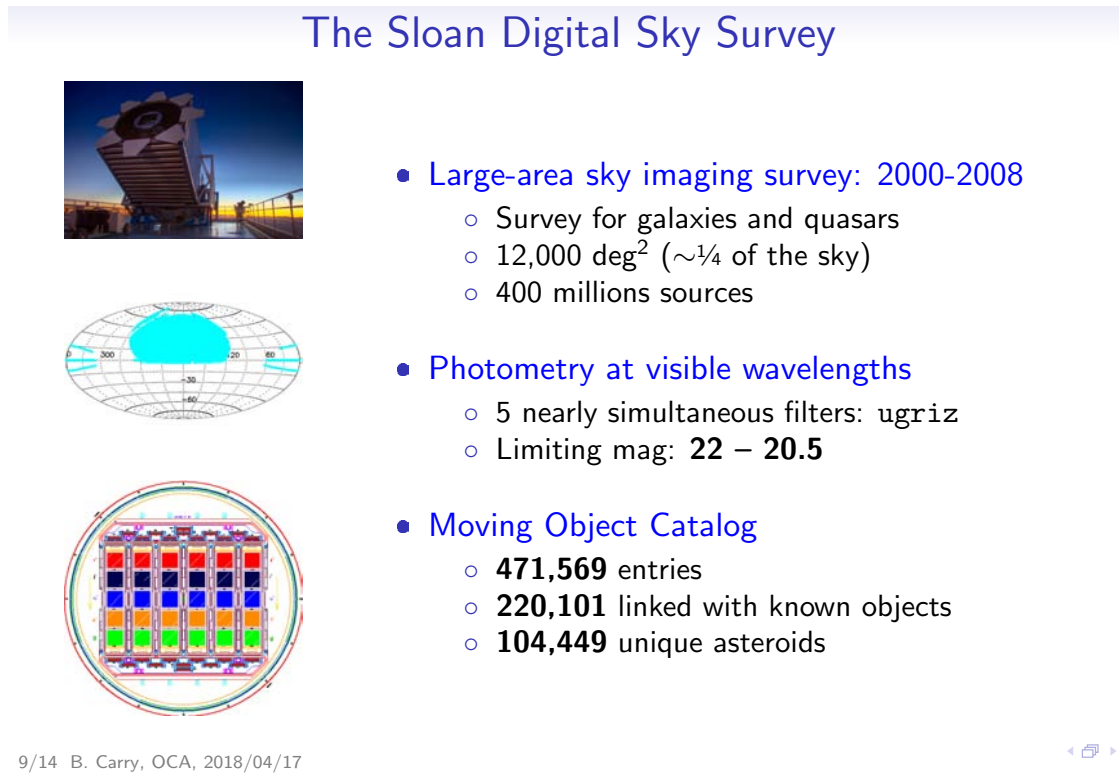

**Figure 5.** From Carry, 2018-ASIME.

**a. Is there any evidence that the orbit of an asteroid provides information on its composition? was Q10. [revisit]**

Yes, but orbits only give *statistical inference* on spectral type because the dynamics are highly chaotic, so they cannot alone be relied upon to identify targets (Green, 2018, Graps, et al, 2016-ASIME).

Dynamical models, such as (Granvik, et al, 2016), can provide an additional hint on the number of NEOs of different sizes. With the estimates of relative contributions to the NEO population from different source regions in the main belt, and some trends in appearance of the different main belt groups, it may be possible to make at least some statistical estimates. **[SKG 5]**

Such inferences are also **size-dependent** (Carry, 2018, https://youtu.be/9SX4nFnIN4M?t=5150 ). See chart below from Carry, 2018, where he explains.

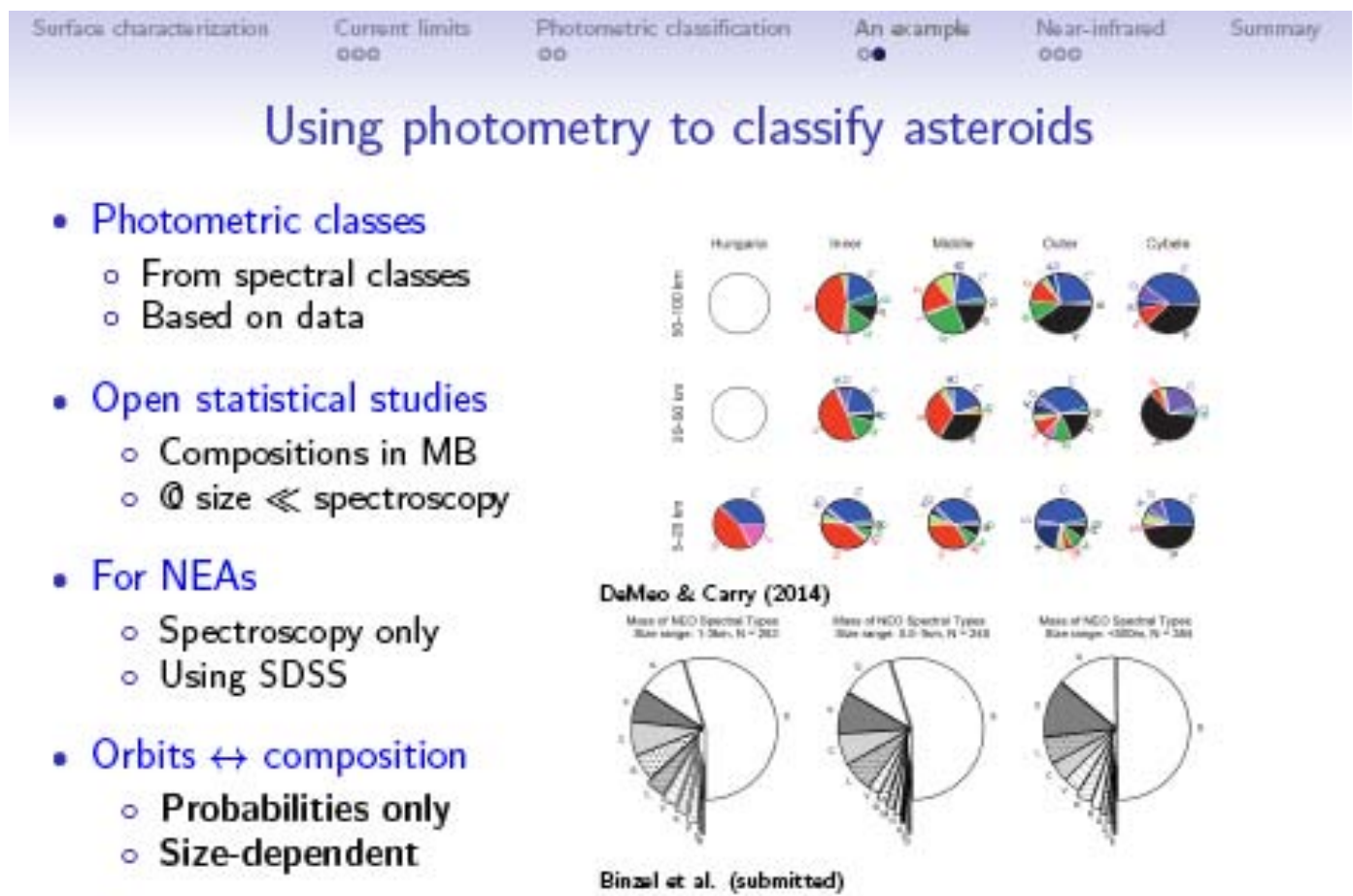

**Figure 6.** From Carry, 2018-ASIME.

For main belt asteroids, this is possible due to *asteroid family* affiliation. The orbit provides

Amara Graps | graps@psi.edu

information of the most likely *source region*.

Although we know that large asteroids are darker in the outer main belt, and brighter in the inner main belt, this "zoning" is not obvious for smaller bodies with NEA sizes **[SKG 5]**, so it's impossible to assess the composition based solely on the orbit (except, maybe for cometary orbits).

However, if we know the *taxonomic type*, then we can estimate the most probable asteroid family that may be at the origin of the object by searching for families close to the most likely source region with the same taxonomic type.

**Taxonomy**: See valuable 15-slide introduction to asteroid taxonomy in de Leon's talk, available in slides and on YouTube (https://youtu.be/7GQgJscCdFE?t=8210)  (deLeon, 2018-ASIME).

***Key Point: Taxonomy*** *does not provide obvious asteroid physics and chemical composition when you find a matched pattern to your data. The physics and chemistry enters when you match the same patterns in the lab. I.e. matching spectra from asteroids to spectra from meteorites. With these links we can identify composition and more.*

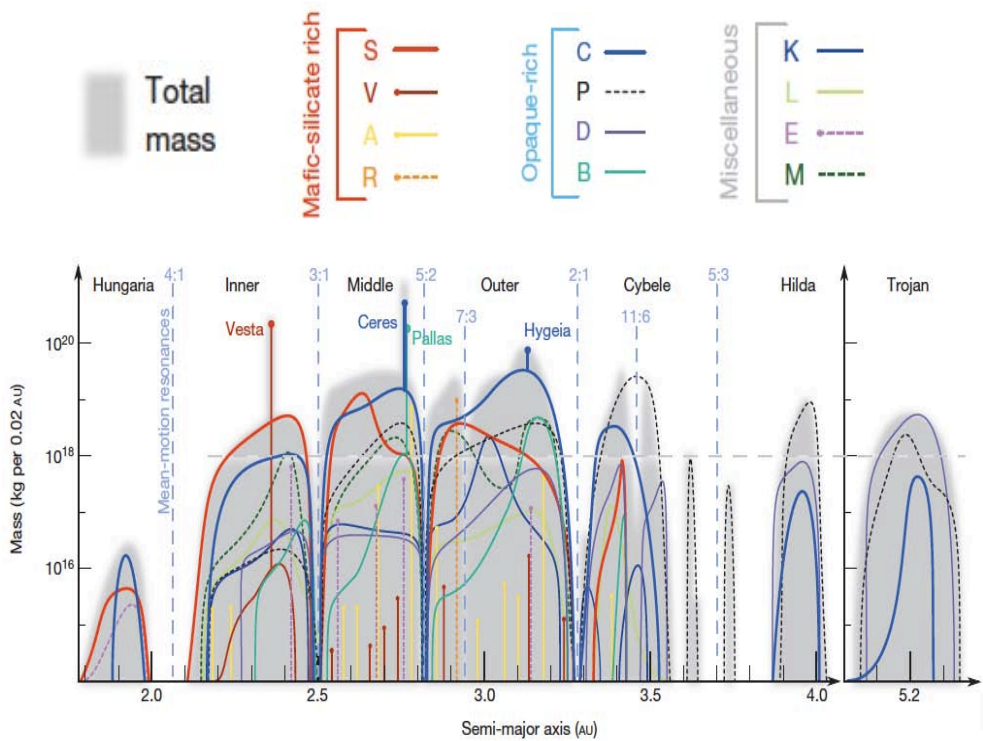

**Figure 7.**  Orbital Distribution. From DeMeo and Carry, 2014 and Green, 2016.

From NEOWISE, the taxonomic distribution of small asteroids:

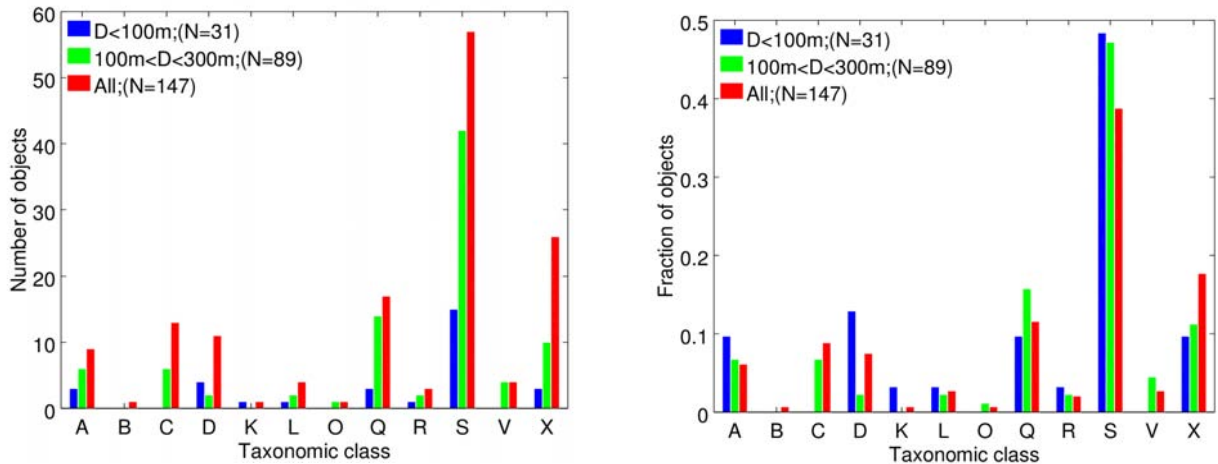

**Figure 8.** NEO Taxonomic Distribution from NEOWISE. From Barucci, 2018-ASIME.

For more data on the **D-type asteroids**, the asteroid type that the Lucy mission will be targeting, see Barucci, (2018).

For NEAs: First, dynamics can be used to infer the most possible origin in the main belt (Bottke et al., 2002; Greenstreet et al., 2012), and then our knowledge about the composition of main belt asteroids can be used to constrain the composition of the particular NEA in question.  **[SKG 5].** The statistical quality of this inference is because:

- Probability of NEA ejection location (no dynamical memory of original source)
- Greater mixing of types at smaller sizes - influence of Yarkovsky effect?

One prominent exception to the NEA (NEO) indeterminism: some NEOs are on typically cometary orbits. So in those cases, it is reasonable to assume that their composition resembles that of cometary nuclei.

**Sample Return Improvements**

Samples retrieved by Hayabusa2 and OSIRIS-REx are expected to bring information about the hydrothermal environment in the parent bodies of Ryugu and Bennu and its effect on minerals and organics (Brunetto and Lantz, 2019).  The results of the sample studies will be used both to support the analysis of retrieved samples and to formulate predictions about the trends that should be measured on asteroid **families** by remote sensing spectroscopy (L1–RS1 in Fig. 2).

Studies from the ground, comparing Ryugu and Bennu, showed similarities and contrasts (Sugita et al., 2019; Lauretta et al., 2019; Binzel, 2019). From the Earth, both bodies displayed relatively neutral-coloured and featureless spectra, with very low reflectivities. However, Ryugu's slightly reddish tinge placed it in the *Cb-class* within the asteroid taxonomy lexicon, whereas Bennu's bluish hue casts it into the *B-class*. (Binzel, 2019).

Ryugu's observed visible spectral type is close to that of the asteroids Eulalia and Polana, which are the parent bodies of C-complex asteroid families in the inner main belt (Sugita, et al., 2019; de Leon, et al., 2018, deLeon, et al, 2018, Campins, et al, 2018). Orbital dynamics calculations demonstrated that the most likely origin of Ryugu is either Eulalia or Polana (Bottke, et al., 2015). Their populations in the inner main belt, from which the largest fraction of Near-Earth asteroids is derived (Bottke et al., 2005), contains many large families, such as Eulalia and Polana, with these spectral characteristics (Morate, et al, 2016).

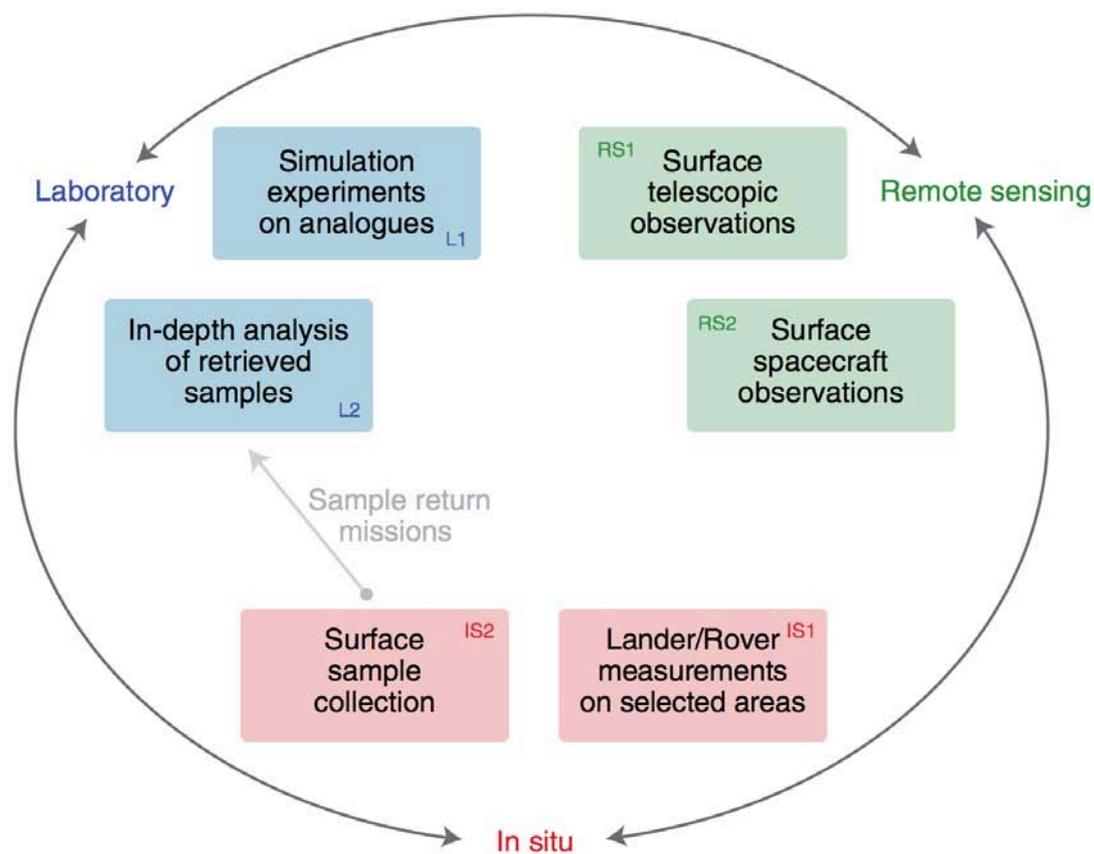

**Figure 9.** Closing the loop thanks to retrieved samples. Sample return missions play a key role in linking laboratory analysis and experiments, remote sensing observations and in situ characterization of Solar System bodies, and in joining efforts among complementary scientific communities. From Brunetto and Lantz (2019).

**b. How can the rate of spectral characterisation of NEOs be increased? It lags far behind discovery rate, especially at smaller sizes (D < 300m).  was Q2. [revisit]**

This question is aimed towards characterizing the population of potential mining targets for which one would ideally want characterization of every single object.

Amara Graps | graps@psi.edu

*Long discussion in the Fitzsimmons discussion-BOX 2.*

<u>A short answer (Green, 2018)</u>
- Requires dedicated follow-up programmes at discovery. **[SKG 2]**

1. Follow-up and characterisation of targets
        - large enough for follow-up spectroscopic observations
        - require physical characterisation – e.g. exclude fast rotators
2. Follow-up at discovery apparition if possible
        - Likely brightest at discovery
        - potential long delay until next good apparition
3. Size range 30 – 100 m
    - Too few at larger sizes

<u>A short answer (Carry, 2018)</u>
- Look in the 'trash' of what hasn't been analyzed in the large sky surveys. Those are the asteroids.

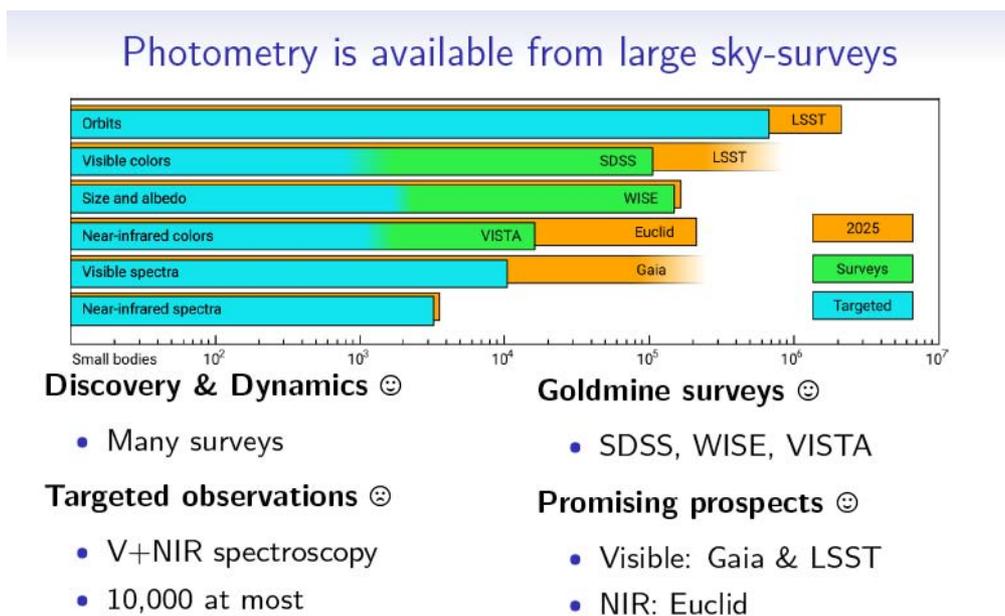

**Figure 10** From (Carry, 2018).

M. Elvis (2018) showed the power of combining the "where" and "what" (orbits and composition) with a single instrument, the .Magellan/PISCO. Question is, **who pays?**

Exploiting **new surveys** can help, but, there are limitations: (Green, 2016)

GAIA  (Operations 2015-2019)
        - Astrometry of all asteroids with V <~20
        - Low-res visible spectra (~150 000 Asteroids)
  LSST (operations 2022-2032)

Amara Graps | graps@psi.edu

- >5m asteroid discoveries
- All-sky survey every few nights in 6 wavebands (0.4-1μm)

**Limitations**:
- Spectrophotometric coverage not optimised for solar system objects (asteroids move)
- No IR coverage

The situation is described in Galache et al., 2015: "The increasing rate of discovery has grown to ~1000/ year as surveys have become more sensitive, by 1 mag every ~7.5 years. However, discoveries of large ($H \leq 22$) NEAs have remained stable at ~365/year over the past decade, at which rate the 2005 US Congressional mandate to find 90% of 140 m NEAs will not be met before 2030 (at least a decade late). Meanwhile, characterization is falling farther behind: Fewer than 10% of NEAs are well characterized in terms of size, rotation periods, and spectral composition, and at the current rates of follow-up it will take about a century to determine them even for the known population. Over 60% of NEAs have an orbital uncertainty parameter, $U \geq 4$, making reacquisition more than a year following discovery difficult; for $H > 22$ this fraction is over 90%."

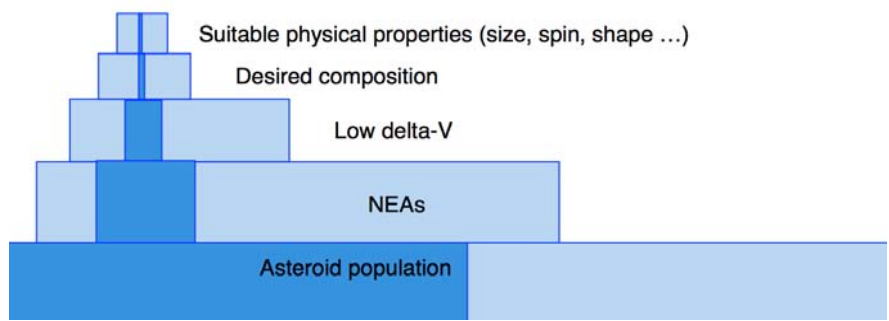

**Figure 11**. A schematic of the small number of accessible and suitable NEAs for asteroid mining missions. The top-level vertical line couldn't be drawn thin enough. (Green, 2016).

From Galache et al., 2015:
"We show that for characterization to keep pace with discovery would require: quick (within days) visible spectroscopy with a dedicated ≥2 m telescope; long- arc (months) astrometry, to be used also for phase curves, with a ≥4 m telescope; and fast-cadence ( < min) light curves obtained rapidly (within days) with a ≥4 m telescope. [...] For the already-known large ($H \leq 22$) NEAs that tend to return to similar brightness, subsequent-apparition spectroscopy, astrometry, and photometry could be done with 1–2 m telescopes"

To characterize at the discovery rate only optical (0.4-1 micron) spectra/colors can be used as the near-IR sky background is large and variable, limiting observations to V~<20. The optical sky is dark, so V=22-23 is possible and it makes more sense to observe in the visible wavelengths. This task will still need large (4+ meter, even 8-10 meter) telescopes. Time can be purchased on many of these, although it's not called that.

NEOs require rapid, "target of opportunity" scheduling (within a few hours, or otherwise certainly within the same night). Such scheduling is not in place at most large telescopes. **[SKG 2]** Rapid response with a dedicated telescope is needed because NEOs, especially the small NEOs, are usually brightest at the time of discovery; that is the only time when they are bright enough to be observed (see "NEO Sizes" in the previous section). The longer the time period after discovery, when follow-up observations are conducted, the larger one's telescope usually needs to be. That constraint can quickly limit available options.

High quality photometric calibration would also be needed for this approach, but to some extent, this will be provided once the Pan-STARRS1 survey releases its catalog of the northern 3/4ths of the total sky. Note that the discovery rate will grow, possibly greatly, once LSST is observing >2021.

While finding/funding a dedicated telescope is critical for search/discover/follow-up for NEOs [**SKG 2**], <u>a more efficient observational strategy needs to be found</u> rather than just finding a way to apply the brute force approach of doing full IR/Vis spectral characterizations of every single NEO, with broadband/narrowband photometry as proxies for spectroscopy.

<u>One observational strategy</u>
Determining the object's spectral slopes would still be useful and can be measured with broadband filters. If possible, narrow-band filters could potentially give you even more information at the expense of not being able to characterize fainter objects. However, the target object's rotation is problematic, when using broadband colors, because one usually observes through one filter at a time, and then subtracts the magnitudes in each filter to derive colors. However, for a rapidly rotating object (which many small asteroids are), the asteroid itself may change magnitude significantly between observations just from rotation.

So for handling rotation, cameras attached to the telescope that obtain observations in multiple filters simultaneously could help with the fast rotating asteroids. These exist but are not widely used. Significant blocks of time on one or more large-ish telescopes (~2m+) with multi-color cameras available could help at least determine spectral slopes for asteroids at a much higher rate than is possible now.

<u>Another observational strategy:</u> A step-by-step approach as in planetary defence (PD) seems useful:
- discovery (present and future PD resources)
- initial orbit determination (here: low delta-v filtering)
- tracking to refine orbit before losing it (same as PD)
- rotation period
- broadband colours
- selected objects: narrowband colours selected objects: spectroscopy.

This avoids wasting large-telescope time on objects that are unlikely to be useful. Variability by rotation can be detected by interleaving b/w exposures (clear filter) with colour filter exposures, possibly even compensated while exposure time is short with respect to the rotation period.

Amara Graps | graps@psi.edu

**Box 2. 1st Day Summary Discussion (Fitzsimmons, 2018)**

**Alan Fitzsimmons:** *How to increase discovery of the small water-bearing asteroids?*

Buy/Build a telescope/orbital observatory? When there is significant capital and maintenance cost?

Or do buy time from existing telescopes (for example, what NEOShield2 does) and then be prepared to share the data? (because it is taxpayer-funded (public) data.)

S/N versus spectral resolution. Filter optimisation? Optical spectroscopy only?

Or is spectrosocopy necessary?

Can you "Triage" ? I.e. First : Photometry (for ex. PSCOPE on Magellan telescope) and then narrow it to potential C class and then do spectroscopy?

Good in *theory* but not in *practice*, because: Many are only able to do spectroscopy in the discovery apparition (1 week or less)

What about Optimization of filters? Apply spectroscopy 0.5-1.0 micron, to differentiate between C class and S class asteroids ?

**Thimo Grundmann**: Alot of asteroid work has a history of using obsolete and underused telescopes. There are a lot of 1.0 m telescopes already working observing asteroids. That would be cost-efficient. That would provide a base amount of public data, so that companies can use that data with minimal cost. There are already examples of such usage with geological data, for example, that mapped by the USGS, which is then used by the oil companies.

**Martin Elvis**: There are large telescopes that are privately owned so that you don't necessarily have to make the data public.

Plus large class telescopes are shifting in obsolesence. As the largest telescopes are being built on the ground, the next larger telescopes are becoming obsolete. For example, now the 4-meter to 8-meter size telescopes have availability. So there are opportunities for accessing the smaller telescopes for NEO observations.

**Alan Fitzsimmons**: Yes, but that shift to obsolescence takes about 10 years. For example the ELT is ONE telescope, and the VLT are FOUR telescopes, which are still in active use. The 8-meter telescopes won't be available that quickly.

**Benoit Carry**: I do agree that going into the archive is opening to objects that we may not care about, there is valuable data existing for free, already there. For example, in the archives we found for NEOSHIELD2 150 NEA spectra right away.

**Alan Fitzsimmons**: I think it depends what you're looking for. If you're looking for *colors,* some surveys are useless. But if you're trying to improve an orbit, and recovery of that object, then surveys like PANSTARRS is helpful.

[**H.-K. Moon:** gives good overview of KMTNet, 3 identical 1.6 m telescope stationed at Chile, Australia, S. Africa. See Moon, 2018 presentation for details.]

**Alan Fitzsimmons**: 2 min exposure: NEA will be trailed. Excellent photometry on those objects. That's a really nice facility.

It won't be enough to catch up the discovery rate though.
I'd like to remind people that in 4 years time, LSST will be online and which will increase the discovery rate by five times. Majority of objects will be so faint that only LSST will be able to follow them up.

Amara Graps | graps@psi.edu

## c. Is there any evidence that the <u>shape</u> of an asteroid provides information on its composition? was Q9. [revisit]

- No, not really. (Green, 2018)

We learn about its evolution, however (Green, 2016). The shape may be a weak constraint for extreme objects (fast spinners) or for small NEOs when YORP process, spin up, would cause primary to shed mass and form a binary system (Reddy).

For information asteroid shape determination, See 2016-ASIME White Paper: (Green, 2016, Graps et al., 2016). Answer in more detail.

The only place where shape of an asteroid seems to be modified (especially for small NEOs) is by the YORP process, spin up, where the primary would shed mass and form a binary system. I (Reddy) have spectrally characterized a large fraction of YORP binaries in NEO population and I see no evidence for a compositional preference for binary formation. On the other hand, it would be interesting to see if a metallic object can form a YORP binary. I think it comes down to rubble pile vs. monolith and so far we are seeing evidence for rubble piles down to meter-sized objects.

Other processes contribute to the shape, such as impacts. The impact response of an asteroid depends on its material composition and porosity. Therefore, the morphology of impact features, that can sometimes contribute to the global shape (e.g. Mathilde, Eros), can tell us something about the composition (a metallic object will show something very different from a carbonaceous one, for instance).

Another clue to the composition by impacts is the secondary ejecta. Secondary craters are formed by re-impacting slow co-orbital ejecta (Murdoch, 2016). Such secondary ejecta provides constraints on the ejecta velocity field and knowledge of source crater material properties (Nayak and Asphaug, 2016)

However, the morphology of impact features can only be seen in-situ. Remotely, combined shape, size and spin information allow us to determine whether cohesion is needed to make the asteroid stable in its configuration. It does not say much about the chemical / mineralogical composition, but can tell us whether it needs to be composed of material with some degree of cohesion to be stable with its size, shape and spin (for an assumed density).

To acquire information on the internal *structure* (not exactly composition) it is useful to evaluate surface structures such as lineaments, crater shapes, crater ejecta, boulder existence / distribution, and mass wasting features.

They could also provide information on material strength, cohesion, porosity etc. both for the asteroid regolith and interior. The size dependent occurrence of such structures is still poorly known. Such surface morphology information gained right before any mining event could support the planning of the activity (spatial properties, sequential order of steps, especially in-situ analysis). The following two examples indicate the concept.

Largest under-degraded crater
One technique to understand the mechanical properties before any drilling takes place is to note the size of the largest under-degraded crater (Murdoch, 2016, Murdoch et al, 2015). It gives information on the mechanical properties of the body, via estimates of the attenuation of stress energy; see Asphaug, (2008). Weak asteroids may survive collisions that would shatter and disperse monolithic solids as they can dissipate and absorb energy better.

Grabens or Linear depressions
While asteroids are covered by loose fragmental debris, the surfaces also exhibit grabens or linear depressions (Murdoch, 2016, Murdoch et al, 2015). It is possible that these fractures are evidence of competent rock below the asteroid regolith. It has been suggested that they result from stresses from large impact events, which have refocused and caused fracture far from the crater (Fujiwara and Asada, 1983; Asphaug et al., 1996), or that they are due to thermal stresses (Dombard and Freed, 2002) and/or body stresses induced by changes in spin. However, faulting can occur even in a granular matrix when it is cohesive relative to the applied stress.

Not about composition. We can remove the shape model uncertainty by a closer look, and get definite composition by in-situ analysis and sample return. See talk: Grundmann et al, 2018.

> **d. What conditions would permit the presence of free water ice on an NEO (e.g., on an extinct comet), and what would be the best way to detect it remotely? was Q6. [revisit]**

*Short Answer and Discussion in the Fitzsimmons summary –BOX 3.*

**Short Answer.**
There must be very special circumstances for such to exist (See Green, 2018 Wrap-up). Highly eccentric orbits, for example. If the ice is present, it is not detectable from remote sensing. The special circumstances are the following. (Rivkin, 2018, 2016).

Due to the NEO proximity to Sun, the temperature $T_{max}$ > 350 K, hence underline{no surface water} is possible; since it should sublimate very quickly. Contrived polar cold spots are unviable.

- Ice: Almost certainly not
- Ice stability *very* T-dependent
  - Need T ~ 110 K for surface stability
  - Need surface T ~ 140 K for shallow burial stability
  - Need surface T ~ 170 K for km-scale burial stability

- Temperature way too high for NEOs
  - 7% albedo at 1 AU: $T_{avg} \sim 280$ K
  - 7% albedo at 1.5 AU: $T_{avg} \sim 230$ K
  - Sub-solar T much higher

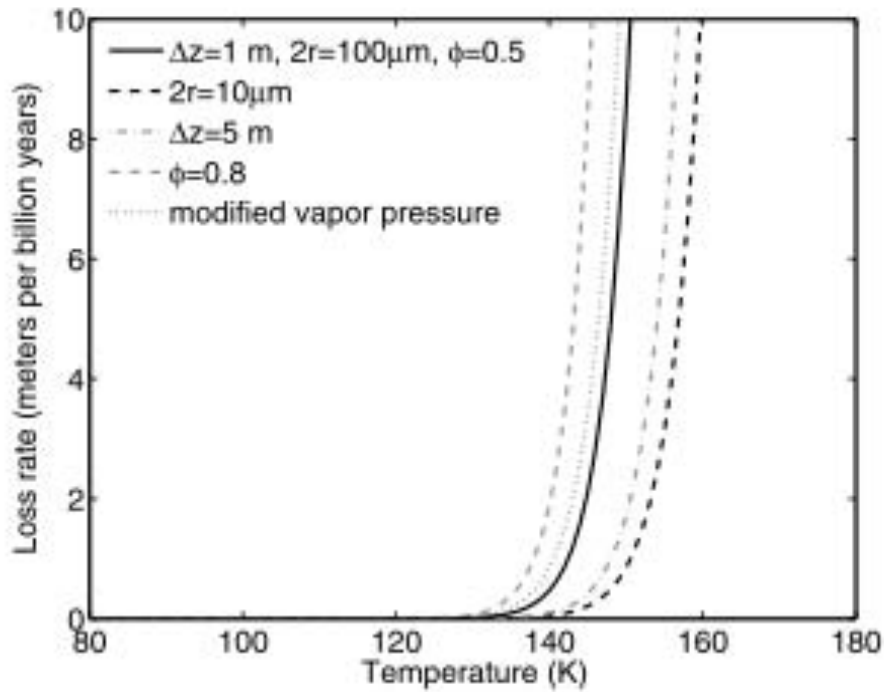

Fig. 1.—Theoretical loss rates for ice buried by material of thickness $\Delta z$ with grain radius $r$ and porosity $\phi$. The solid black line is a reference case with parameters listed in the first legend entry. For the other line types, the legend indicates the parameter which differs from the reference case. The calculations use the equilibrium vapor pressure for crystalline ice, but one graph with a modified vapor pressure, from Bryson et al. (1974), is also shown, where $\ln p_s = -\Delta H/(RT) + 21.7$ (in torr) and $\Delta H = 11.4$ kcal mol$^{-1}$ for 132–153 K.

Figure 12. Modeling by Schorghofer, 2008.

Amara Graps | graps@psi.edu

> **Box 3. 1st Day Summary Discussion (Fitzsimmons, 2018)**
>
> **Alan Fitzsimmons**: None! (Ask Andy). Go to a normal comet or a Main-Belt Comet if you're that interested. Or go to Ceres.
>
> **Humberto Campins:** There could be ice on Bennu.
>
> **Andy Rivkin:** There could be some, but is it worth getting? I don't think that there would be an economically viable amount of ice at the poles of Bennu.
>
> **Joel Sercel:** We looked at it and decided that it is not economically viable. Asteroid mining companies should not count on ice in low-delta-v NEOs.
>
> **Simon Green:** We could all imagine a scenario where the asteroid obliquity is close to the ecliptic, and there is a deep hole where ice could sit. However, asteroids, unlike the Moon and Mercury, don't retain their geometry. Both in terms of distance from the Sun and their obliquity is constantly changing. You won't find ice until you are at the asteroid anyway. To have both of those conditions support life is not likely, so the answer is no.
>
> **Andy Rivkin:** If there is an impact onto the icy body.. we don't know how it got that way.
>
> Obviously we would look for it, but I don't think you would find it.
>
> **Joel Sercel:** This is largely an academic discussion. It is only economically viable for asteroid mining companies to support asteroids that are in Earthlike orbits.

### The possibility of subsurface ice

While asteroid subsurface ice being present for long periods (life of the Solar System) in the Main Belt at 2-3 AU is possible, such subsurface ice on NEOs wouldn't have that longevity.

In the Main Belt: depending on the thermal properties (e.g., composition, porosity, grain sizes, etc.) of the dust, ice can remain for a surprisingly long period of time, potentially longer than the relatively short dynamical lifetimes of NEOs, but the NEO temperature is critical, see below. Schorghofer, (2008) demonstrated that buried ice on spherical bodies, within the top few meters of the surface, orbiting 2-3 AU from the Sun, can survive ~$10^9$ years.

For NEOs: (Rivkin, 2016, 2018)
Note that: At 300 K, ice retreats at about 1 km per My, even under a 1 km dry layer.

- Ice retreats ~330 m/My at 280 K beneath 1-km dry layer
- Slows to 3 m/My at 230 K
- 60 m/My at 230 K beneath 50-m layer

So you'd need a thermal environment where the body is rarely at high temperatures, such as in a high eccentricity orbit, and it would likely need to be large enough to still retain any ice. Furthermore, the changing obliquities and low gravity of NEOs means no permanently shadowed regions near the poles.

Therefore, the combination of traits needed to retain ice for NEOs: (Rivkin, 2016)
- Very short time spent close to Sun: eccentric orbit
- Not much time spent at high temperature: dynamically young
- Sufficiently-deep insulating layer: lag deposit

In other words, you need an actual comet (and not an extinct one).

<u>Surface impacts to expose the ice</u>

Let's say there is more evidence that subsurface ice exists on a NEO. In principle, an impact by a micrometeorite (or something even larger) on one of these objects could briefly expose near-surface ice, which would then quickly sublimate and potentially cause an outburst of comet-like activity.  It might be hard to detect in the plume, because dust scattering dominates, but an optical imaging survey looking for such outbursts (and using techniques capable of detecting extremely low levels of activity) could do it. This approach would not detect \*all\* NEOs with near-surface ice, but could be a way of detecting some of them.

<u>Tagish Lake meteorite pores as evidence of subsurface ice</u>
As a nugget for thought, there is some meteorite evidence of ice surviving to down to the surface of the Earth. The Tagish Lake meteorite likely had pores with at least some ice when it was collected. The Tagish samples that were curated frozen appear to retain some ice.

> **e. While low-to-medium resolution spectroscopy in the 0.4-4.0 micron range is the best way to obtain a taxonomic classification of an asteroid, is it possible to obtain similar results using colour photometry? was Q3.**

*Yes!*

<u>Short Answer</u>.
- Useful for initial identification of primitive types
- Careful choice of filters can indicate presence of some features
- Benefits of easy observation (and large surveys)
- Fast and efficient and which will get you results right away (see Carry 2018).
- Machine-learning is the ultimate way to go. We don't want to invent a new taxonomy; we want to use existing taxonomies so that we can benefit from the decades of laboratory work. (Carry, 2018)<u>Longer Answer, See Discussion in **Box 2**</u>.

Amara Graps | graps@psi.edu

Longest Answer, see the detailed explanation in presentation (Carry, 2018, https://youtu.be/9SX4nFnIN4M?t=4951).

For first step of taxonomic classification of asteroid types, Carry suggests using photometry, it will get you far, and then you can refine. For example, trying to identify a rare D type in the main asteroid belt had a 50% success rate with using photometry of large surveys.

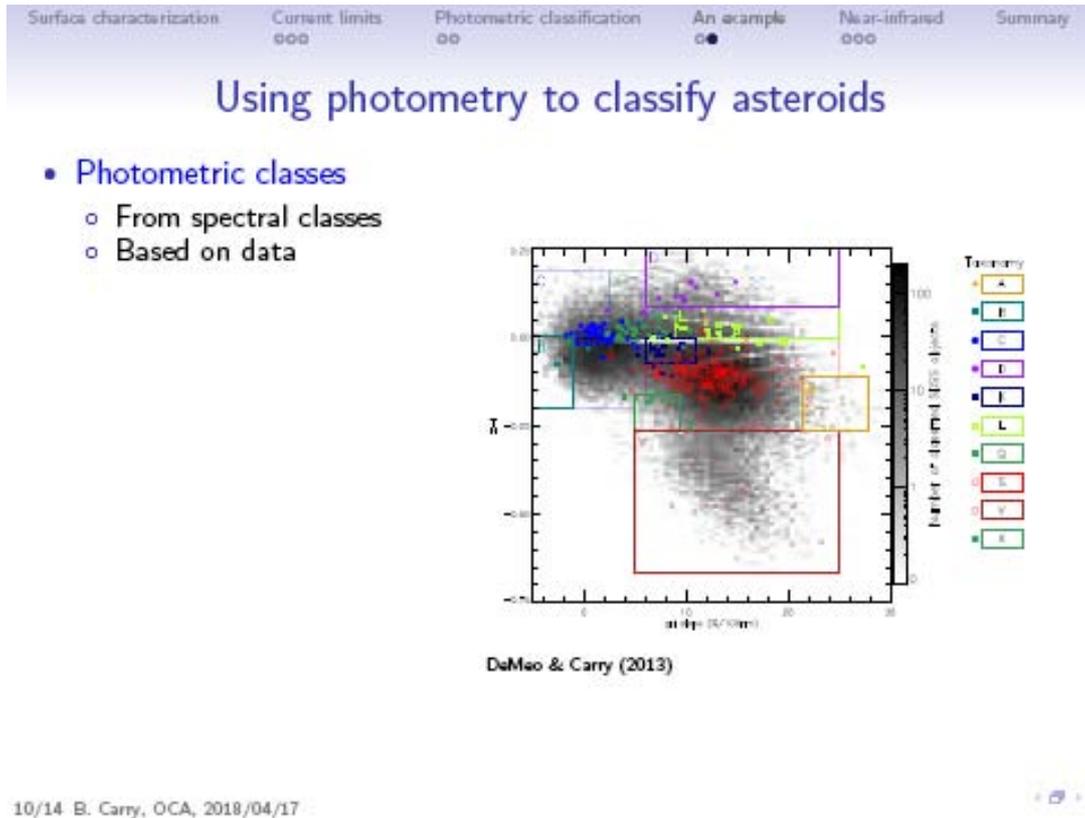

**Figure 13**. A faster way to get you in the vicinity of taxonomic classification. From (Carry, 2018).

Here, from two years ago: white is discovery, grey is just spectroscopy, red is SDSS, not designed for asteroids with a citizen science activity: increase of 600% for discovery (Carry, 2018-ASIME, https://youtu.be/9SX4nFnIN4M)

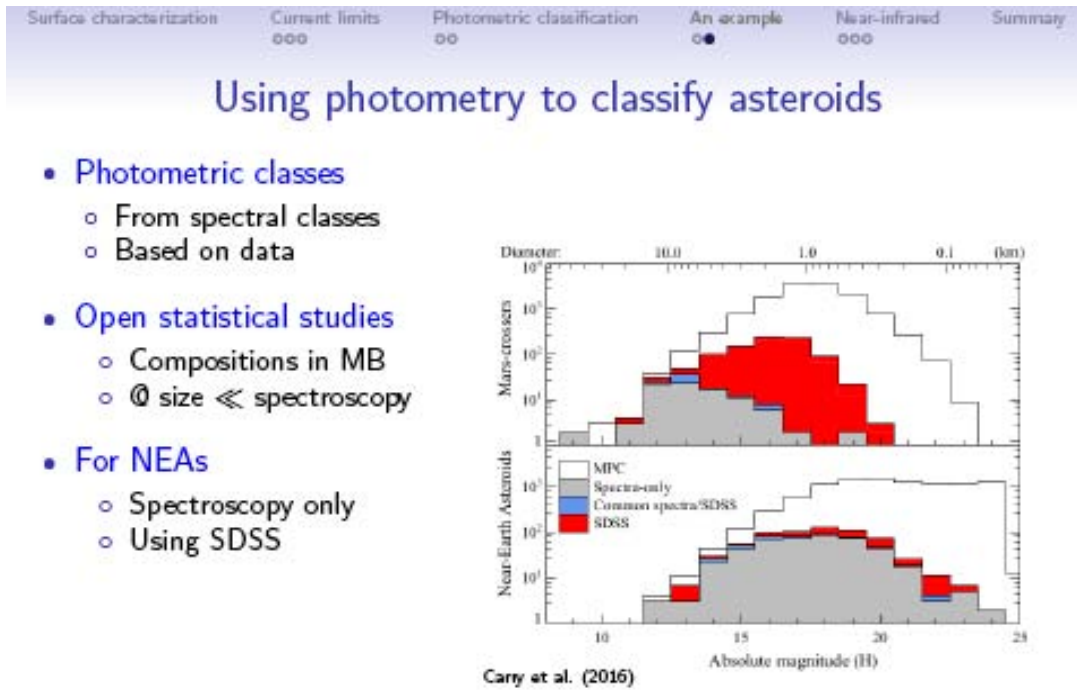

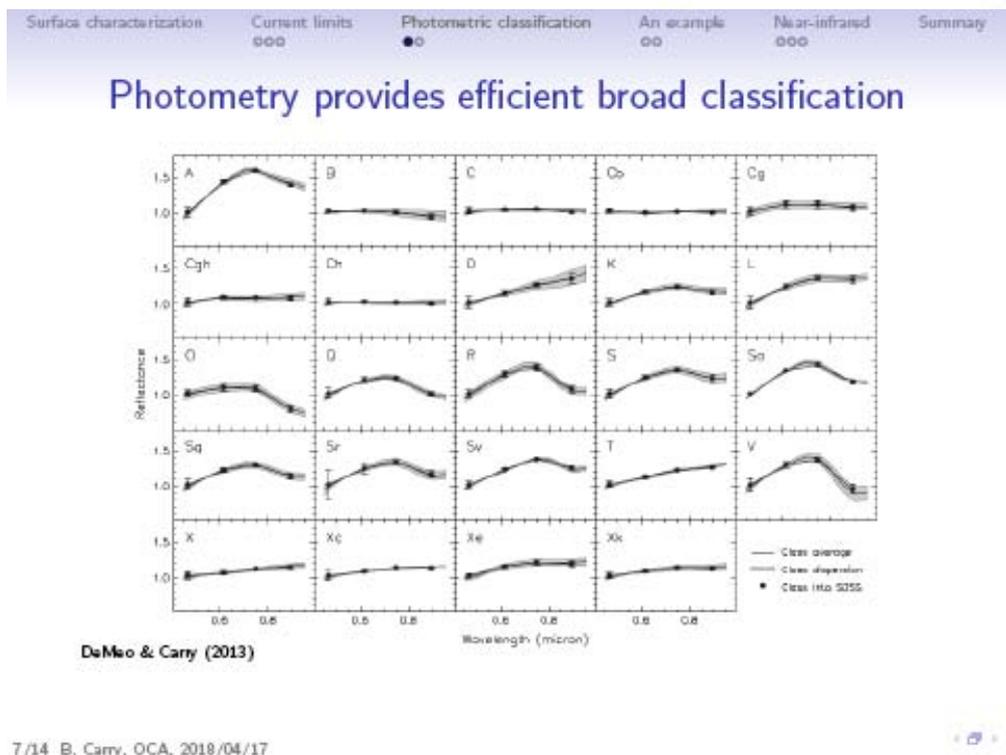

**Figure 14**. Top: Bottom: from (Carry, 2018).

We characterise surfaces from reflected spectra (Reflectance) and the asteroid albedo. Then compare those reflectances and albedos with meteorites (V+NIR+MIR).

It would be wise to keep in mind the lessons from 21 Lutetia (Barucci, 2018-ASIME), with its surface inhomogeneities. Lutetia is an old object (with a surface age of 3.5 Gy) with a primitive chondrite crust and a possible partial differentiation (Weiss et al. 2012) with a metallic core. Lutetia's surface appears to be a mixture of "incompatible'' types of materials: carbonaceous chondrite (dominant) and enstatite chondrite (in minor percentage).

This particular taxonomy, the Bus-DeMeo system, has a wide use in the planetary community, but there are others. See talk and presentation by de Leon, 2018-ASIME.

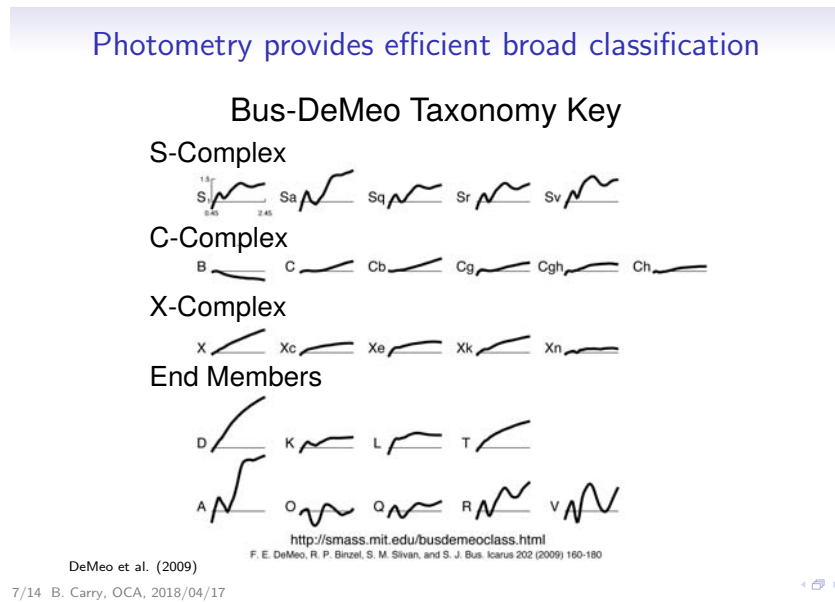

**Figure 15**. Bus-DeMeo Taxonomy System from (de Leon, 2018-ASIME).

Fitzsimmons says in the discussion of the Green Wrap-Up that there are a number of instruments available on at least 4-meter class telescopes that are a combined camera with low-resolution spectragraph, to support both kinds of measurements in the discovery apparition of the NEA. So you would use the color photometry to narrow down the object type to be the type you want, then do spectroscopy right away to characterize the NEA much better.

> **f. How can the water absorption feature at 3.1$\mu$m be best used as an indicator of hydrated minerals on carbonaceous asteroids? What additional measurement would further increase the quality or fidelity of the measurement? Was Q5. [revisit]**

From Green (2018) ASIME wrap-up: "It's an odd question., essentially it is asking: what's better than the best method from the remote observational viewpoint, so let's explore the question a little bit. "

For asteroid spectra , the 3.1 micron data exists mostly for Main Belt Asteroids and larger objects; very few small NEAs!

Of the small numbers of primitive asteroids with hydrated minerals:

- Smaller targets have S/N too low?
- Near 1 AU (where best low delta-v targets reside), the <u>region is dominated by thermal emission</u>
- Space-based system doesn't help because the thermal emission from the asteroid is dominant.

Rivkin partially disagrees, and says that we don't know yet how big a problem the thermal emission is and whether OVIRS? and NIRS3 measurements are possible or not. Both Green and Rivkin suggested a high phase observation could solve the problem, with Green saying that the phase needs to be 'really high phase': greater than 90 degrees. But then there is a problem for the small objects with never having enough signal. Conclusion: Don't rely on it. See conversation here: https://youtu.be/9SX4nFnIN4M?t=26712

**The 0.7-micron feature easiest to detect.**
- Note that it is a weak feature in the optical
- $Fe_2^+ Fe^+$ feature – always associated with 3.1 micron OH
- Only seen in fraction of cases where 3.1 micron feature is seen
- This feature can be observed with a wide range of telescopes

Of the 0.7 mm absorption band as a proxy for hydration, see more
from de León et al. (2018-ASIME) https://youtu.be/7GQgJscCdFE?t=9325
and from Rivkin (2018-ASIME) https://youtu.be/7GQgJscCdFE?t=23019

# 4. Primitive asteroids

Ground-based observations of the absorption feature associated to phyllosilicates in the 3-mm region is almost impossible, due to the Earth's atmosphere

We can use the presence of the 0.7 mm absorption band as a proxy for hydration

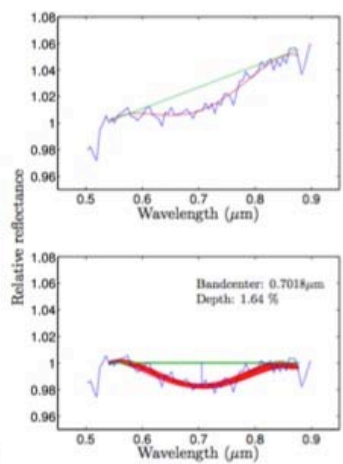

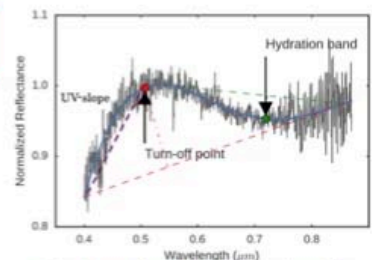

Fig 4. Example of parametrization of the visible spectra (De Prá et al. 2017)

Parametrization:
- ✓ Turn-off point
- ✓ 0.7 absorption depth
- ✓ UV-slope

For the slopes we use the spectral gradient S' expressed in %/[0.1 μm]

16 April, 2018

**Figure 16**. Using a 0.7-micron absorption as a proxy for water from (de Leon, 2018-ASIME).

**Other diagnostics:**
- 0.9 micron – no benefit over 0.7 microns
- 2.4 micron X-OH – very weak
- ~6.5 micron mineral bound $H_2O$ emissions – requires spacecraft
- 8-12 micron silicate/phyllosilicate CF – requires large telescope
    with thermal IR spectrometer or spacecraft and large/close targets

The presentations: Barucci (2018-ASIME), Rivkin (2018-ASIME), and Campins (2018-ASIME) go into more details about water-rich asteroids and NEOs, their formation and detection.

Another technique: Moon's 3D color taxonomy correlations (Moon, 2018-ASIME) are able to identify the water-rich C-type taxonomic class (DeMeo taxonomy) in SDSS data.

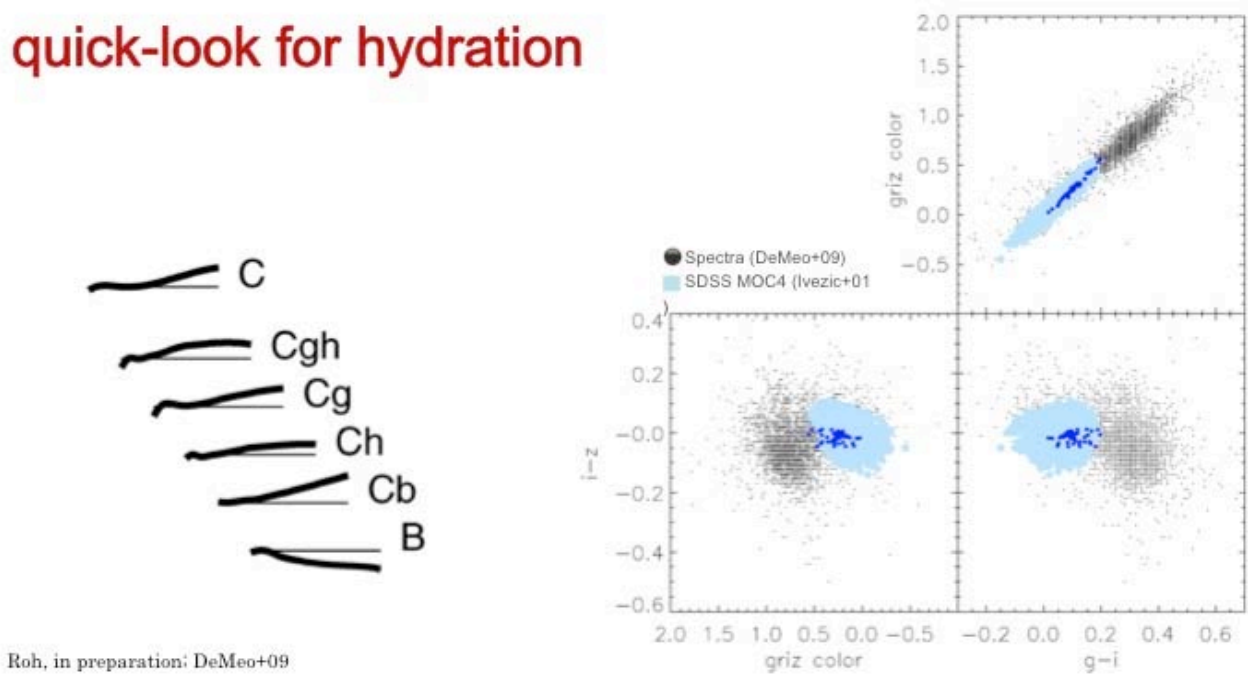

**Figure 17**. Using a 3D taxonomy correlation method to match for water: H-K Moon, 2018-ASIME.

H-K Moon continues this discussion with their source regions in in presentation (Moon, 2018-ASIME. https://youtu.be/9SX4nFnIN4M?t=4102 )

---

### g. What highest value telescopic composition/characterisation studies are not being pursued for lack of funding or perceived low priority from space agencies? Was Q11. [revisit]

Green (2019) in Wrap-up: Any and ALL of them! We aren't getting the support from space agencies beyond planetary defense purposes, for dedicated NEA follow-up, so will likely have to do it 'piece-meal'.

Grundmann (2018-ASIME) is more specific and provides these examples: solar sailing (e.g. Gossamer type), suitable nanolanders (e.g. MASCOT type), and specific MNR trajectory studies.

Short and Long answer in ASIME 2016 White Paper (Graps et al., 2016) and fundamental to **SKG 2**.

A short answer: Green, 2016.

- Priority for mitigation funding is still on making NEO discoveries
  - Spectroscopic follow-up requires dedicated telescope **[SKG 2]**
  - Observations made at discovery maximises S/N
  - Multi-band imaging for fainter targets

- Competition should *not* be with astronomical facilities
  - One-at-time targets are hard to compete with multi-object programmes
  - Light curves require extended observations
  - Objectives are not (primarily) science driven

- Large surveys can help
  - Astronomical (non-asteroid) surveys often limited value for NEAs
    - Limiting magnitudes to bright; non-optimal observing strategy
  - Expect some improvement from LSST and Gaia

Ground-based facilities

Currently, for ground-based instruments the discovery of new NEOs has larger priority over dynamical or physical characterization. It is clearly difficult to obtain telescope time at large facilities for performing time-demanding surveys (for instance, by spectroscopy) of a large number of targets. Many "new" or known NEOs are not followed up at all (some of them- actually "many" for small sizes ~140m - can even be "lost" as too few astrometric observations follow discoveries). The situation is not going to improve with the advent of LSST and an increasing rate of discoveries - simply the fraction of characterized NEOs will decrease. To solve this problem, dedicated telescopes are probably the most efficient approach. .
**[SKG 2]**

Amara Graps | graps@psi.edu

Ground-based facilities: Dedicated, Robotic Telescopes

Dedicated (possibly, fully robotized) facilities on the ground could be very performing on several aspects (light curves for shape determination, visible-NIR low-resolution spectra, astrometry).
Ground-based facilities: VNIR, Visual

Some suggested that a large scale VNIR survey of the NEO population would be helpful. Something like 1 year dedicated survey on NASA IRTF or UKIRT. If we can produce a SMASS-like survey for small NEOs in the NIR that would be helpful for answering a number of questions here. Others disagreed by pointing out that NIR can't go fainter than V~20. Because most NEOs are now discovered at V~20-21 and then fade, NIR is (for now) impractical. Instead it was suggested that we have to use Visual band (0.4-1.0 micron) only. It will depend on what exactly the required rate of false positives/false negatives from the mining side is. Visible wavelengths (the "0.7-μm band") could work if you don't care about false negatives (though there are details to that as well).

Ground-based facilities: Radar
The best method against losing asteroids, particularly relatively small and faint ones near Earth which quickly fade out of telescopic view again, is immediate radar observation on the discovery conjunction. This gives an orbit accuracy comparable to a visual telescope recovery at the next conjunction. But this requires facilities like Goldstone or Arecibo in size, which would have to operate in close contact responding to the visual wide-field surveys' output - very expensive

Ground-based facilities: Multiple filters (colors), instead of spectra and Network of small telescopes
A large-scale color characterization survey for NEOs capable of rapid response to new discoveries and using a camera capable of observing in multiple filters simultaneously could be pretty useful. Colors would not be as good as spectra, but can at least give you spectral slopes, and simultaneous multi-filter observations will remove uncertainties due to rotational brightness variations. As mentioned above, rapid response will mean that a follow-up telescope may not need to be as large as would otherwise be needed since NEOs, especially small ones, are typically brightest at the time of discovery, and fade quickly as their distance from Earth increases. A network of such telescopes (even just 2 or 3) would be even more useful to allow follow-up of a larger number of NEO discoveries than would be possible from just a single telescope from which not all newly discovered NEOs will be observable for various reasons.

Ground-based facilities: Harnessing Underused Small Telescopes

There are a number of 1-meter or larger telescopes around that were in scientific observatories but became obsolete for 'big science' or were replaced by more advanced or larger ones. A survey of such under-used, mothballed or become-museum telescopes would be useful to see whether they could be converted for some type of spectroscopy. All of the early Planetary Defence surveys were

Amara Graps | graps@psi.edu

based on decommissioned light buckets, but in that case, they were wide-angle (in astronomers' terms) designs e.g. made for satellite tracking. Now the more typical astronomy telescopes would be useful which have a large aperture but also small field of view because this better suppresses background light from the atmosphere or unresolved objects. Also, top class amateur telescopes now have passed the 1 m aperture mark.

Dedicated Space-Based Facility/ies

Telescopes in space can have the best observing conditions and access wavelengths that are precluded to ground-based instruments. A study not pursued for lack of funding is the investigation of satellites constellation dedicated to NEOs observation from space. Having such system in space would be surely more expensive and more risky than just relying on a ground-based network, but it would also provide the consistent advantages of being free from the limitations imposed by weather conditions, by the day/night cycles and by the scattered light into the atmosphere. Some food for thoughts: the system could be deployed in a single dawn-dusk SSO, observing in the anti-sun direction perpendicular to the polar plane, to exploit good and stable illumination conditions and to have similar dynamical perturbations for all the satellites (that would make the constellation more stable). The scanning strategy would have to be defined accordingly to a trade-off between tracking and surveillance and it would lead to the selection of the minimum number of satellites. The constellation could be increased along time, but it seems reasonable to have something already functional with 4 spacecraft, thus within a single launch with a heavy carrier. Complexity would be surely high, but the constellation would also have an operational flexibility impossible to be achieved with ground-based systems.

Space-based (single) telescope example: AsteroidFinder (at DLR)

AsteroidFinder was an advanced design, ~150 kg class spacecraft, using an electron-multiplied (EM) CCD and a 25-cm-class telescope. The design was carried through to Phase B (and a B2 and B+) but then discontinued. If flown, the telescope would have pointed towards the Sun to discover and track asteroids at solar elongations >30°. The scientific purpose was to discover more asteroids Interior to Earth's Orbit (IEO) with the goal to at least double the known population, i.e. find another 10 or so in 1 year. It would also have caught hundreds of Aten NEAs >100 m, and a lot of Apollos. Plan B was a conventional CCD with image stabilization. Using a dawn-dusk Sun-synchronous (DDSSO) orbit sped up the discovery rate and improved tracking of discovered or known objects. Summary, such a design could be achieved by one agile <200 kg spacecraft. One could implement two or three, based on the same design with different sunshade shapes.

# II. Asteroid - Meteorite Links

**Importance of same observation geometry:**
Is the reflectance measurement of meteorites in the lab, where you don't usually make backscattering measurements <u>at the same observation geometry</u> as that on the asteroid?
Are the <u>thermal environments</u> in the two the same (thermal gradient critically affects the spectrum, see Donaldson-Hanna, 2018-ASIME)? There are <u>grain-size dependencies</u>, too.
See P. Beck's 2018-ASIME presentation for a detailed discussion on similar observation geometries.

Surprising news of Tagish Lake CM chondrites: Meteorite appears to be thermally dehydrated (Bonal, 2018-ASIME). This is only one of two D-type analogues in the entire meteorite collection.

Six asteroid types, ~2/3 of the mass of the asteroid belt, are absent from our meteorite collections (Barucci, 2018-ASIME). Carbonaceous chondrites are only 5% of meteorites collection. Meteorites contain up to 22% of water.

> **h. What is the state of the art regarding matching meteorite spectra to asteroid spectra, and matching artificially weathered meteorite spectra to asteroid spectra? was Q22.**

> **i. Is anyone working on software that models how weathering affects meteorite spectra, to then attempt to match asteroid spectra to this modelled weathered meteorite spectra? was Q23.**

Two suggestions: one in hardware and one in software and then a discussion of laboratory state-of-the-art.

<u>Hardware</u>: The Complex Irradiation Facility at DLR (Grundmann et al, 2018-ASIME) provides these services to irradiate dry materials to the conditions of interplanetary space.

- ultra-high vacuum ~$10^{-10}$ mbar
- $p^+$ & $e^-$ @ 1...100 keV, 1 nA...100 μA
- UV/vis/NIR 40...2500 nm
- sample @ -193...+450°C

<u>Software</u>: Here is a correlation method by (Moon, 2018-ASIME) that applies broadband visible colors (g-i, i-z, griz) in "3D" to match space weathered meteorite taxonomic classes; for example the DeMeo classes for ordinary chondrites (L class) and S-type asteroids. Space weathering vector follows the trend, see the next figure.

Amara Graps | graps@psi.edu

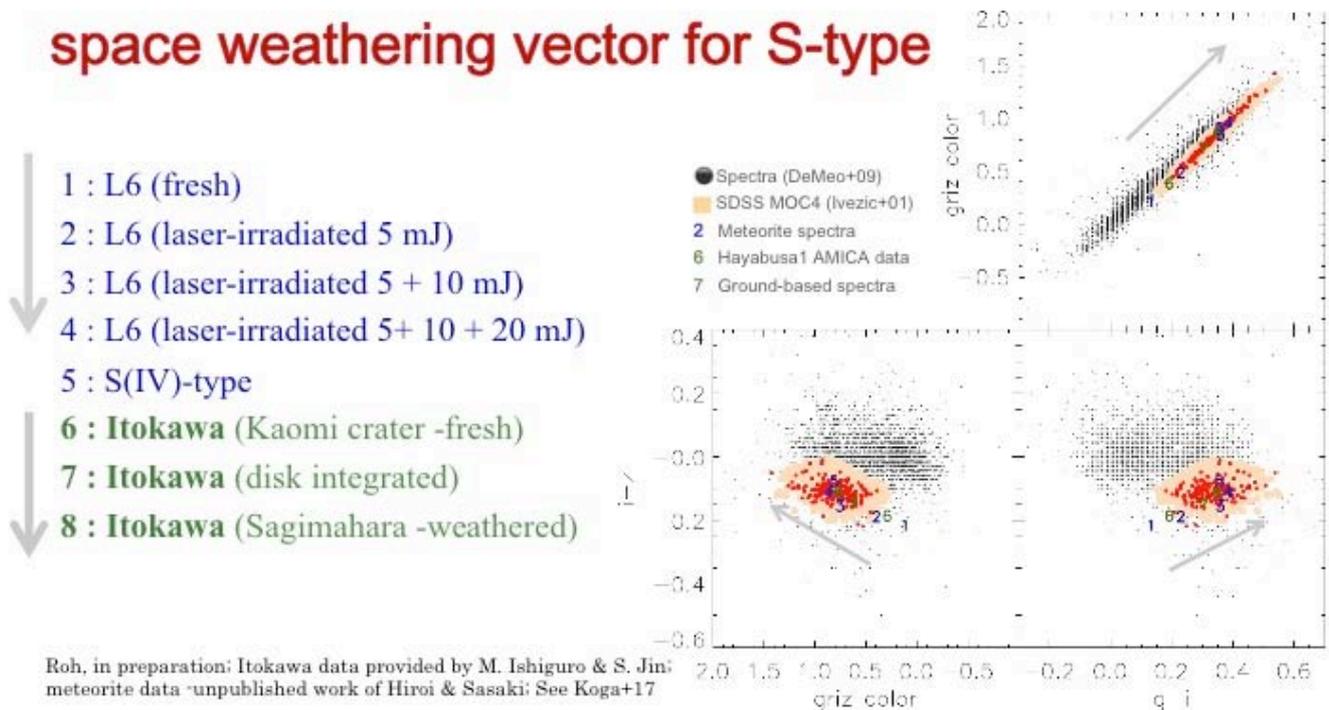

**Figure 18**. Using a 3D taxonomy correlation method to match space-weathered Ordinary Chondrites (L-type) with S-type asteroids: H-K Moon, 2018-ASIME.

From the ASIME 2018 presentation by Donaldson-Hanna (2018-ASIME), and below in Fig. 19, about meteorite and asteroid spectral matching, know that: Physical mixtures behave differently under simulated asteroid conditions than chondritic meteorites. These differences are not subtle, and important for parameter fitting of asteroid and meteorite spectra.

In particular, both meteorite spectra show a spectral contrast increase near 10 microns as well as an increase in the spectral contrast of the transparency feature. These differences are likely related to two important differences between the analogues and the meteorites. The first is the analogue mixtures were of significantly higher albedo than the meteorites, which affects the thermal gradient induced in the sample. And second, the analog mixtures contain terrestrial phases and synthesized materials that are supposed to best represent meteorite phases, but are likely different.

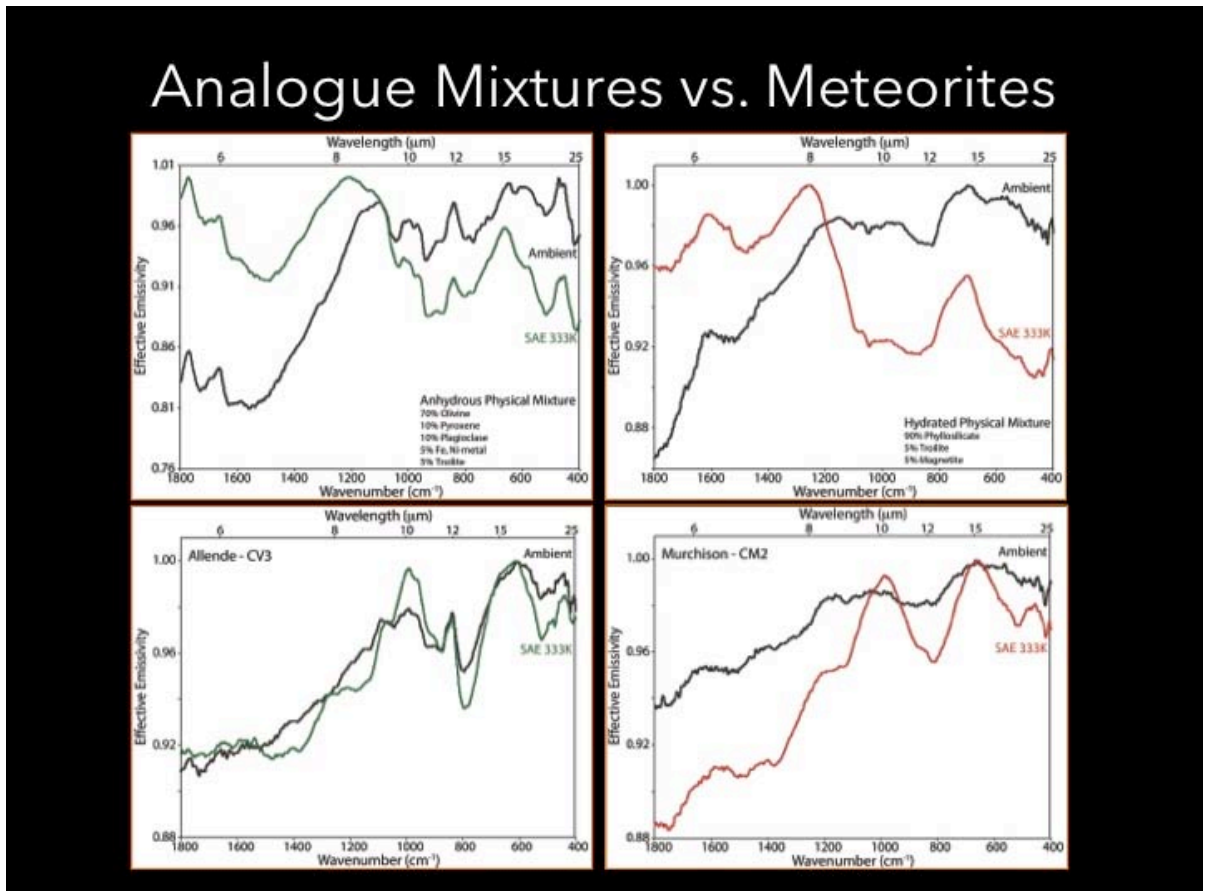

**Figure 19**: Analogue Mixtures vs. Meteorites by Donaldson-Hanna (2018-ASIME). Under Earth-like or ambient conditions, we see good similarity between the meteorites and their analogues. However, under simulated asteroid conditions we see that analogues behave closer to what we observe in pure mineral spectra (e.g. the CF shifts to higher frequencies and an increase in the spectral contrast is observed between the CF and the RB), but the meteorites show different effects.

For other state-of-the-art methods, see deLeon presentation, (2018-ASIME) for matching meteorite spectra to asteroid spectra, both classical methods and variations on the classical methods.

> **j. How well understood are the processes of space weathering, and can we tell what the original state of the surface was, based on the current state? was Q15. [revisit]**

Space weathering is one of the processes studied in the last 20 years that shows that asteroids are not the pristine objects that we thought they were. It is a still an active area of research and but – some- corrections are being applied to understand the asteroid's original surface spectra, especially for S-type asteroids. The classic reviews of space weathering are those by Hapke (2001), Clark et al., (2002), and Chapman, (2004), which were reviewed and updated in a chapter in the Asteroids IV book by Brunetto et al., (2015) and more recently, Pieters and Noble (2016).

Since the publication of the Asteroids IV chapter, among new results are useful laboratory experiments by Kuhlman et al., (2015), from Dawn space mission data to Vesta, by Blewett et al.,

Amara Graps | graps@psi.edu

(2015), and recent work on C-type asteroids by Kaluna et al, (2016).

From the Brunetto et al., (2015) Review in the Asteroid IV chapter, from the observational point of view, slope trends from large surveys and in particular of young asteroid families have confirmed that solar wind is the main source of rapid ($10^4$–$10^6$ yr) weathering. The author states that the diversity and mechanisms of asteroid solar wind (SW) need to be investigated further. There are some indications that Q-type asteroids may essentially be unweathered S-type asteroids (e.g., Binzel et al. 2010; Willman et al. 2010), but the process is much less well understood for C-type asteroids. He points to a particular situation that VIS-NIR slope variations are still controversial for dark, C-type asteroids, but there's been new work since then.

Kaluna et al., (2016) present visible spectroscopic and albedo data of the 2.3 Gyr old Themis family and the <10 Myr old Beagle sub-family. The slope and albedo variations between these two families indicate C-complex asteroids become redder and darker in response to space weathering. Their observations of Themis family members confirm previously observed trends where phyllosilicate absorption features are less common among small diameter objects. Similar trends in the albedos of large (>15 km) and small (615 km) Themis members suggest these phyllosilicate feature and albedo trends result from regolith variations as a function of diameter. Observations of the Beagle asteroids show a small, but notable fraction of members with phyllosilicate features. The presence of phyllosilicates and the dynamical association of the main-belt comet 133P/Elst-Pizarro with the Beagle family imply the Beagle parent body was a heterogeneous mixture of ice and aqueously altered minerals. The apparent mineralogical differences between the Veritas family and the Themis and Beagle families highlight the importance of accounting for mineralogy when interpreting space weathering trends across the broad population of C-complex asteroids.

From Dawn space mission data to Vesta, Blewett et al., (2015), and Vesta presents some open questions about V-type and S-type asteroid space weathering. The results show that as the regolith matures, it becomes darker and bluer (i.e., the 438-nm/555-nm ratio increases). This is the spectral trend predicted for addition of carbonaceous chondrite material to basalts by exogenic mixing, whereas lunar style space weathering (LSSW) should lower the reflectance but cause spectra to become redder (lower 438-nm/555-nm ratio) as npFe0 accumulates. The lack of obvious LSSW on Vesta continues to be a puzzle, because V-type and S-type asteroids are redder than powdered samples of their respective meteorite analogs [SKG 1]. Such reddening is consistent with space weathering by solar-wind ions. While the presence of the bluish CC component in Vesta's surface could partially mask the reddening effects of npFe0 accumulation, meteorite evidence suggests that npFe0 production on Vesta is indeed very low.

In 2015, Kuhlman demonstrated that the process of solar wind implantation on airless surfaces like asteroid surfaces can be reproduced in the laboratory. Kuhlman shot hydrogen atoms at solar wind speeds into tiny, polished samples of the common Solar System mineral orthopyroxene that had been placed on top of a silicon wafer. She then examined the compositional changes in the outer 20 nanometers of the implanted orthopyroxene using a scanning transmission electron microscope (STEM), and for the first time discovered the particles of iron beginning to form.

YunZhao Wu presented lunar space weathering science in his ASIME 2018 presentation (Wu, 2018-ASIME and https://youtu.be/9SX4nFnIN4M?t=1625 )

Amara Graps | graps@psi.edu

Space weathering affects the spectral slope in this way in the Clementine data: topsoils have a shallower slope, while freshly escavated craters show a deeper slope. The lunar top soil spectral signature in the visible and near-infrared wavelengths, such that with the 415/750 nm ratio of Clementine data, decreases with increasing topsoil maturity, i.e., older is bluer. This finding is consistent with the ultraviolet observations for the Moon (Denevi et al., 2014) and asteroids (Hendrix and Vilas, 2006), and in the presentation by Barucci, (2018-ASIME). Implications: remote sensing spectra are most suitable for learning about the regolith, while knowledge about the mineralogy should be from freshly exposed rocks.

Grundmann said that the Hayabusa 2 small body rover: MASCOTs can likely scrape the surface to expose fresher materials in the mm…cm depth range. We will learn more of what it discovered soon. See talk: Grundemann et al, 2018.

> **k. Among the scientific community, what is the current confidence that spectral class informs bulk composition, given space weathering and the results from recent missions connecting asteroids with certain spectral classes to known meteorite types? was Q21.**

P. Michel: I don't think the current confidence is high.

We have a vast knowledge of **heterogeneity** at microscopic scales in each meteorite in our meteorite collections. (Green, 2018, Wrap-Up: https://youtu.be/9SX4nFnIN4M?t=27136)

For example:
- Almahatta-Sitta has a mixture of evolved and primitive material
- This is at microscopic scales. Is the macroscopic regolith mix uniform?
- Such heterogeneity complicates the linking of composition and spectra

We think we have a broad understanding of many (but not all) links between particular classes of meteorites and particular asteroids. The Hayabusa 1 samples match predictions from Itokawa spectra. Space weathering is understood and verified for S-type asteroids.

For dark C-type asteroids, we have a growing understanding of the weathering effects from lab analogues. See Beck and Bonal, and Bonal and Beck presentations: 2018-ASIME.

Sum over all of the inclusions
Note that when we take a spectrum, we are taking a measurement of all of the tiny components combined. An asteroid spectra is similarly averaged over all of the boulders in that measurement.

Sample Biases
We should also understand that those we pick up on the ground bias our meteorite collection. The most primitive, most fragile, meteors never made it to the ground. Lesson: If one goes to the asteroid and takes a sample, then that ambiguity is removed. This is one of the biases we expect to remove with the sample return missions (Hayabusa2, OSIRIS-REx). See Green, 2018 ASIME Wrap-up for this discussion: https://youtu.be/9SX4nFnIN4M?t=27423

<u>Temperature Dependence of Hydration Signature</u>
Heated CM chondrites show a darkening and the disappearance of 0.7-µm band. Water and the 3-µm absorption are still there (in a weaker signature). High-velocity impacts on CM chondrites dehydrate the sample, resulting in a disappearance of the optical signatures of hydration. This can perhaps explain the lack of hydration among C-type NEAs.

Correlations using the 3D technique (H-K Moon, 2018-ASIME) with asteroid types and meteorite classes:

- Broadband visible colors (g-i, i-z, griz) are confirmed to be diagnostic tool that can be easily reproduced for taxonomy.
- 3D representation of the colors resolves taxonomic degeneracy to separate seven distinct taxonomy classes.
- However, systematic spectroscopic survey and lab experiments are required to further improve the taxonomy as it is evolving**.**
- It's cheap & simple to use for **science** (evolution), **risk assessment** (bulk density), and exploration & **utilization** (REE & water)

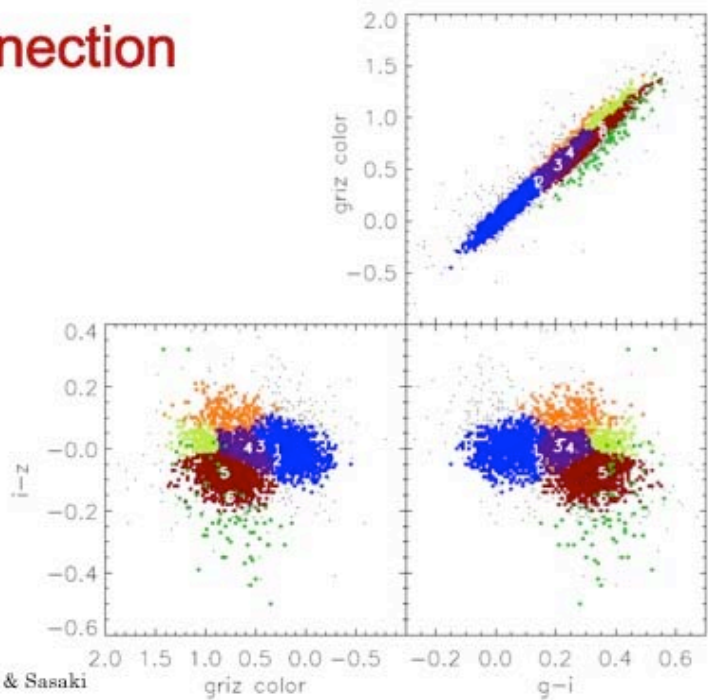

**Figure 20**. Using a 3D taxonomy correlation method to match a variety of chondrites with the asteroid Psyche: H-K Moon, 2018-ASIME.

Space weathering can be included in this spectral modeling method. See where de Leon describes it in her presentation: https://youtu.be/7GQgJscCdFE?t=9227

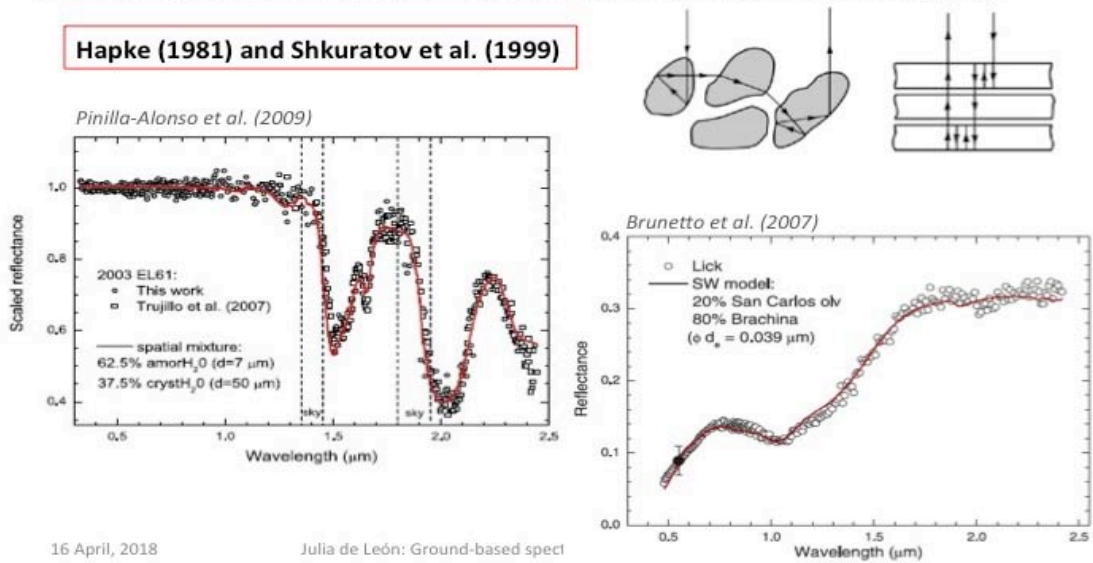

**Figure 21**. Meteorite-Asteroid Spectral Matching by de Leon, 2018-ASIME.

We need samples from primitive asteroids for this kind of C-type asteroid-meteorite linking; **Hayabusa 2 and OSIRIS-REx will return samples soon.** As of April 2019, studies from the two missions, from their in-orbit measurements have appeared in the research journals.

> **I. Could we develop asteroid material simulants based on meteorites; how well do meteorites represent the NEO population, especially at larger (D > 10m) sizes? was Q14. [revisit]**

From Green (2018-ASIME): This is looking at the wrong question. We can probably make simulants that <u>match the meteorites that we have</u>. But those meteorites may not match what is in the asteroid belt, simply due to selection effects. (See sample bias in answer to Q21 above). An important meteorite selection effects is the atmospheric entry or the meteor, which mitigates against low strength, primitive objects.

To explain: Let's say we want these simulants to test an instrument. We can only say that the instrument tests a hypothetical regolith and not the true regolith, because there is missing asteroid information not included in the asteroid simulant.

Amara Graps | graps@psi.edu

# III. In-situ measurements

Following from Green (2018) Wrap-Up. https://youtu.be/9SX4nFnIN4M?t=27514
Now assume that we are at the asteroid.

What we know in April 2019:
- Asteroid morphology has been studied for a few asteroids.
- We have identified some surface mobility such as: seismic shaking, landslips, and ponding.
- Of asteroid ***physical properties***:
    - There have been no lander, controlled measurements <u>yet</u>; the sample returns so far are touch & go
    - Limited regolith properties (grain size, cohesion), which have been inferred from mid-IR observations/thermal models, mostly, so far.
    - potential indirect data from DART/Hera?

All of the above is changing rapidly presently with Hayabusa2 and OSIRIS-REx.

For example:  MASCOT is carrying sophisticated instrumentation with added soil mechanics measurements on the lander, plus the Impactor experiment on Hayabusa2 will return a great deal of soil mechanics data. On OSIRIS-REx, for example, there are images taken during touchdown and sample-head data on the soil.

Other measurements of in-situ regolith physical properties we must include:
- electrostatic dust is critically important (patched charge model). See informal collection of notes in Graps, (2018).
- granular physics is weird and very important in microgravity (M. Elvis). Note: At Johnson Space Flight Center there is a new lab called Hermes, which is going to test granular physics at the Space Station. Experiments can run 6mo to a year. (P. Abell).

We will need to combine all of the above (Hayabusa2 and OSIRIS-REx and the other measurements) with physical laboratory data.  The surface measurements from the reconnaissance missions are needed to constrain physical analogues.

**Mission Issues-**
If the miners are going to benefit from science missions, we have two asteroid missions running, and DART and MMX in the pipeline (2025) and Lucy (to add to water abundance information in the D asteroid population).  The timescales for science exploration missions is long. E.g. missions with small landers have been proposed, but not so far selected.

Amara Graps | graps@psi.edu

**m. Technically, and scientifically, how does spectroscopy of an asteroid at short (km range) distances differ from spectroscopy with ground---based telescopes? was Q4.**

Addressed by S. Green in his ASIME 2018 Wrap-up (Green, 2018-ASIME) and here:
https://youtu.be/9SX4nFnIN4M?t=28090

o Same limitations of top few microns, thin veneer on the surface.
o Spatial resolution will allow study of spectral diversity (if present).
o 3-micron feature possible (for low temperature latitudes). We'll have a lot of signal at the terminator.

Grundmann et al (2018-ASIME) answered straightforwardly: Go there and find out for many NEAs and for several classes.

P. Michel says that we will have more information very soon from Hayabusa2 and OSIRIS-REx, the two sample return missions in progress now. We will be able to learn what spectral types are confirmed by remote observations versus what is collected on the ground.

See again the conclusions from P. Beck's 2018-ASIME presentation about the impact of similar observation geometries on outcomes for asteroid observation and their meteorite analogues.

Yunzhao Wu in his ASIME 2018 presentation demonstrates how they can be quite different:

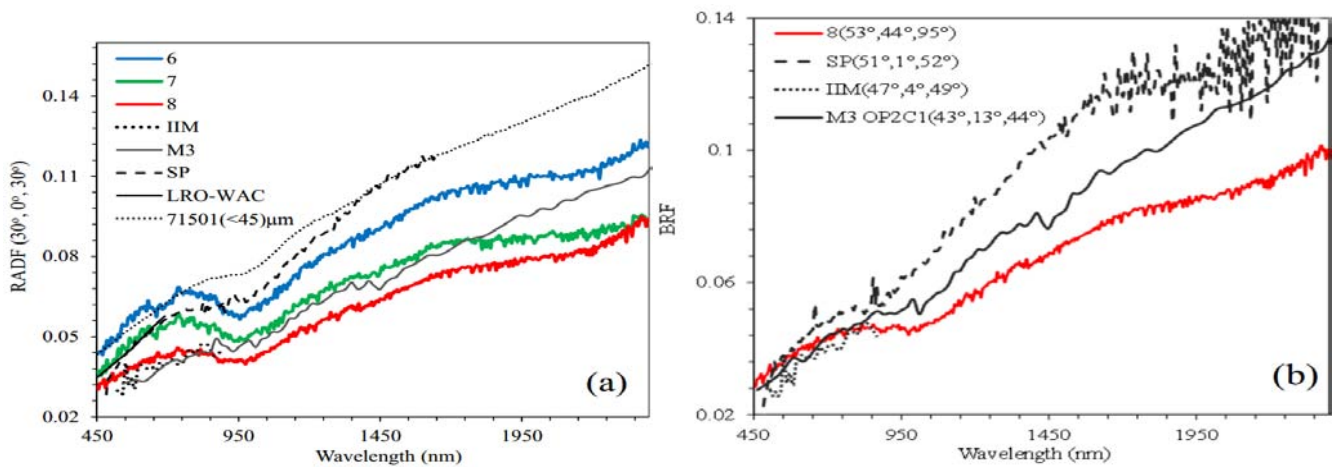

**Figure 22**.: From Yunzhao, 2018-ASIME.

We talked previously about the spatial heterogeneity in micrometeorites.
What about the ***Spatial*** variations in spectra of small NEAs?

o It's real.
o Does it reflect differences on the surfaces of the asteroid, or do they reflect different observing conditions on the asteroid?

Amara Graps | graps@psi.edu

- o In-situ measurements  before Hayabusa 2 and OSIRIS-REx don't show (significant) heterogeneity, but those previous small asteroids (Eros, Itokawa, Gaspra) are not primitive.
- o Variation may be linked to YORPoid surface mobility?
- o Ryugu/Bennu test?

**n.  What instrumentation should an exploration probe carry in order to establish with 100% confidence that water and/or hydrated minerals are present on an asteroid, and what further instrumentation, if any, would be required to ascertain how much water there is? was Q1, Q8. [revisit]**

Addressed by S. Green in his ASIME 2018 Wrap-up (Green, 2018-ASIME) and here:
https://youtu.be/9SX4nFnIN4M?t=28148

- Can't get 100% confidence without landing.  Can't get 100% confidence with landing either! (Unless you are carrying a very advanced payload).

 Remote sensing: Near IR 3um and mid-IR spectrometers. These will help build confidence.

- o X-ray spectrometry gives elemental abundances (O but not H)
- o Neutron spectrometry for H (water inferred)

If you want to be sure: you need to be on the surface with instruments:  mass spectrometers, GC-MS, LA-MS etc. to measure the composition of the species you want. More confidence means more instruments equals more money. It is a question about how far you would go if you were funding reconnaissance.

Grundmann (2018-ASIME) answer:  It is possible to revisit the asteroid with another sailcraft that has a more appropriate MASCOT aboard or that carries a sample-return lander and capsule. The information gained from spectroscopic instruments at the surface or from 'orbit' can be compared to in-situ analysis and sample return. With a MASCOT, the knowledge can be improved in-situ, with mapping and particularly for small NEAs because such asteroids dominate the catalogs.

**o. How could neutron detection support prospecting activities, and what is the maximum depth at which a neutron detector could detect the presence of water? was Q7. [revisit]**

Addressed by S. Green in his ASIME 2018 Wrap-up and here:
https://youtu.be/9SX4nFnIN4M?t=28230

Amara Graps | graps@psi.edu

There is a problem with signal to noise on small planetary bodies, but you can keep integrating to beat down the noise. The penetration depth for H (water inferred) with neutron spectrometry is about 1 meter (as I understand). Therefore, if you are looking for water, rather than hydrated minerals, I don't think that you can get deep enough with this method.

Some think that radar is the best instrumentation for this task (ASIME 2018 Discussion). Radar should be able to provide info on the subsurface water ice. Green answered that we may not get an unambiguous answer. Others answered that we can get that unambiguous answer.

## p. What proximity observations and measurements would better link remote observations to meteorite studies. was Q18. [revisit]

The link between asteroid taxonomic class and meteorites is a **critical link**. **[SKG 1, 3, 5]** How near is 'proximity'? Green in the ASIME 2018 Wrap-Up addressed this asteroid miners' question: https://youtu.be/9SX4nFnIN4M?t=28409

Green said that the 'holy grail' is sample return. Then we will have the ground truth in the lab and the result will be unambiguous. From the sample returns, we will have:

- o   resolved spectroscopy to identify heterogeneity
- o   lander analysis (composition and physical properties)
- o   returned samples MUCH more valuable.. they provide the 'ground truth'

Here are more details (Graps et al., 2016 in the ASIME 2016 White Paper).

From remote observations, we've described what is needed for linking. See  Figure 23 and the presentations: (Green, 2016; Duffard, 2016, Rivkin, 2016, Delbo, 2016)  for some linking of meteorites to asteroid families.  The **[SKG 5]**  describes more scientific information regarding the modeling of the Main Belt source region of NEAs  for a stronger linking between the two: asteroid taxonomic class and meteorites for gaining composition information.  The answers to ASIME 2016 White Paper Question 1 gives a summary of the remote observational techniques needed to determine composition of the asteroid.  The following space missions will gain 'Ground truth' data between volatile asteroid taxonomic classes and meteorite types:

- •   Hayabusa 2 and MASCOT to C-type Ryugu,
- •   OSIRIS-REx to B-type Bennu,
- •   AIM. Now HERA (with radars) for AIDA mission to Didymos?

One would presume that the asteroid is characterized 'enough' before arrival in the proximity of the asteroid. However, as described in the **2018 ASIME Mission Design Discussion: <u>reconnaissance missions or straight to mining?</u>** (see below) is a complex answer, and dependent on the asteroid mining company's strategy.   The answer to this question also depends on the asteroid mining company's desired <u>asteroid resource</u>. Volatiles? Metals?

See also Answers to ASIME 2016 White Paper Question 32: "Asteroid Mining Strategy: What observations can you make from the ground to avoid going to a target of which you have no interest going?" (Graps et al, 2016)

<u>When spectroscopic studies aren't enough</u>

If you manage to characterize the asteroid spectroscopically, it is not enough to understand the asteroid's surface regolith characteristics for a spacecraft landing **[SKG 3]**. For example, Eros and Itokawa have similar spectra and albedos, if one compares a m^2 patch on each asteroid. However, the two asteroids have very different regolith properties.  See the next Figure.

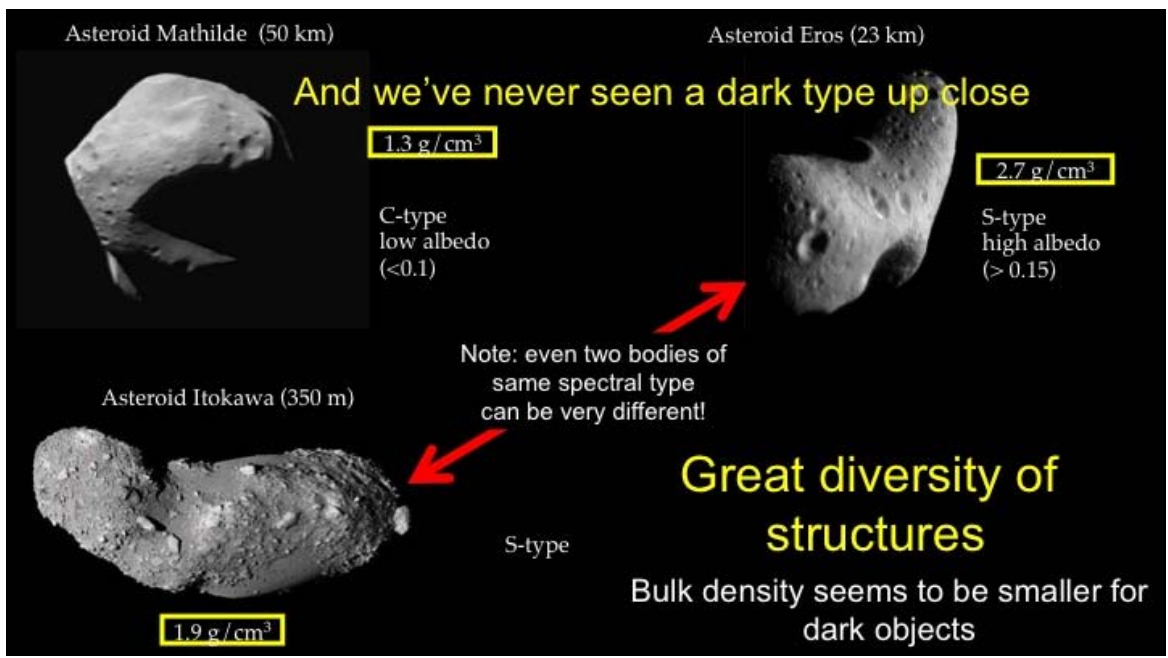

**Figure 23**.: Comment on the scarcity of observations for C-type asteroids and why 'touching' (interacting with the regolith) is important. (Michel, 2016).

The asteroid surface regolith properties can be captured by the <u>thermal inertia parameter</u>. Thermal inertia is the resistance to temperature changes.  Eros has fine regolith size, low thermal inertia.  On the other hand, the Itokawa grain size is the size of tens of cm, very blocky and shows a large thermal inertia. Beware of temperature effects, and compaction effects. Supporting work can be found by Gundlach and Blum, (2013), who calculated the grain sizes for a variety of asteroids from the thermal inertia. In addition, thermal inertia measurements can also help you to weigh the asteroid from remote observations or proximity. Finally, thermal inertia measurements can also be used to identify metal rich NEAs (Drube and Harris, 2016). In that work,  the scientists fit thermal infrared data with a thermal model to identify M-type asteroids while learning about its surface composition, radar albedos, spin rate, roughness of texture, and more.

# IV. Laboratory measurements

Laboratory analyses of asteroid samples and meteorites in the context of asteroid compositions were presented at the ASIME 2018 in the early part of the second day of the workshop. These provided summaries of 1) What we know, 2) deep knowledge of meteorite composition at scales down to nm, including space weathering, and 3) state of the art in simulants.

- o ***Lydie Bonal and Pierre Beck*** <u>Heating processes in primitive asteroids as revealed by the study of organics and hydration of CMs and ungrouped C1/2 chondrites)</u> *Next two videos were not recorded.*
    - ▪ Presentation:<u>http://geophy.uni.lu/users/tonie.vandam/asime-2018/presentations/bonal.pdf.gz</u>

And

- o ***Pierre Beck, Lydie Bonal et al.*** <u>Quantifying hydration from IR signatures of primitive meteorites</u>
    - ▪ Presentation:<u>http://geophy.uni.lu/users/tonie.vandam/asime-2018/presentations/beck.pdf.gz</u>
- o ***Kerri L. Donaldson-Hanna et al.*** <u>Analogue Materials Measured Under Simulated Asteroid Conditions: Insights into the Interpretation of Thermal Infrared Remote Sensing Observations</u>
    - ▪ Video: <u>https://youtu.be/9SX4nFnIN4M</u>  (*only last ½ was recorded*)
    - ▪ Presentation:<u>http://geophy.uni.lu/users/tonie.vandam/asime-2018/presentations/donaldson.pdf.gz</u>
- o ***Wu Yunzhao*** *Key* <u>In-situ spectra from Chang'E-3 and laboratory spectra of meteorites</u>
    - ▪ Video: <u>https://youtu.be/9SX4nFnIN4M?t=15m16s</u>
    - ▪ Presentation: <u>http://geophy.uni.lu/users/tonie.vandam/asime-2018/presentations/wu.pdf.gz</u>

Plus the first day asteroid simulants description:

- o ***Dan Britt and Kevin Cannon*** <u>Simulating asteroid materials with realistic compositions</u>
    - ▪ Video: <u>https://youtu.be/7GQgJscCdFE?t=3h3m50s</u>
    - ▪ Presentation:<u>http://geophy.uni.lu/users/tonie.vandam/asime-2018/presentations/cannon.pdf.gz</u>

> **q. Can regolith simulants be developed that are similar enough to the real thing that experiments would provide accurate results useful to define engineering requirements? was Q13. [revisit]**

Green in the ASIME 2018 Wrap-Up addressed this asteroid miners' question:
<u>https://youtu.be/9SX4nFnIN4M?t=28461</u>

Amara Graps | <u>graps@psi.edu</u>

Yes, and to some level of fidelity, people we know are already making that in the laboratory. For testing the in-situ instruments, chemical simulants are being made and marketed: contact K. Cannon at the 'Exolith Lab' at the University of Central Florida, which acquired in full the simulants formerly produced by Deep Space Industries (DSI). See Cannon's 2018-ASIME presentation, and from 2016: Delbo, 2016, Metzger et al., 2016, Britt et al., 2016. See the Appendix in the ASIME 2016 White paper for a breakdown on the properties that the DSI Simulant considered and included.

We must make sure that we don't follow circular reasoning- that is, that the simulants provide a way to test the asteroid environment, without having the meteorites from which the simulants are based, also match the asteroids.

There is also a need for <u>physical</u> simulants for testing mining processes and the stability of landers/facilities on the asteroid surfaces.

**To address these two needs: chemical and physical, are the samples from the confirmed primitive asteroids from Hayabusa2 and OSIRIS-REx.**

For now, with the exception of meteorites (and their bias), and some information on Itokawa, we have no hard information on the fidelity of simulants.

More from the ASIME 2016 White Paper (Graps, et al, 2016):
Because of the many poorly known factors it is difficult to provide realistic simulant. However as one useful approach, meteorite samples and their powdered (with roughly same thermal inertia like certain measured asteroid surfaces) could be a sample type worth of using during tests. Properties of meteorite powders are a weakly explored topic however, where laboratory tests which provide useful new information that could support instrumental tests.

<u>Case Study: Thermal cracking of meteorites / Generation of asteroid regolith</u>

Delbo and colleagues took Murchison CM2 and an OC meteorite and temperature-cycled the meteorites and calculated what happened to the internal structure. Zooming in one of the slices with x-ray tomography, and analyzing cracks inside of the meteorite revealed the result. After one month they saw visual growth of the crack. This led to the prediction that regolith can be generated by breaking up rocks via thermal cycling (Delbo et al., 2014).

An on-going NASA-funded asteroid simulants development project recently identified 65 characteristics of asteroid materials: However, the project decided that it would be far too difficult to create a simulant with all 65 of those characteristics. Instead, the attendees of a simulants workshop decided to focus on just 12 or 13 of those characteristics. See Appendix 1.

Amara Graps | graps@psi.edu

Similarly, the lunar program identified about 30 properties and decided to simulate only two or three (particle size distribution, a measure of mechanical strength, and to a much lesser degree the mineralogy). The creators of the lunar soil simulant made no effort to replicate the important magnetic and electrostatic properties, for example, which are crucial for developing many technologies. It was decided that it would be too expensive to create a simulant that replicates most or all of the properties. This led to problems in the lunar program because simulant users often mis-used them, assuming that if it is a simulant, and then it can be used in place of lunar soil for any type of activity.

The correct way to use simulants involves considering how they differ from the actual space materials, designing tests that take those differences into account, and interpreting the results in light of those difference. Simulations to extrapolate into the space environment should follow the experiment, or subsequent (more expensive) experiments that use real space materials such as lunar soil brought back in the Apollo or Lunakhod programs. If these steps are followed intelligently, then simulants can be very helpful.

<u>Considerations of making simulants sufficiently high fidelity to be useful.</u>

- First, there is little in the published literature about bulk composition of meteorites **[SKG 1]**, but there is some, and that enables us to develop simulants with all the primary mineralogy (everything more abundant than some arbitrary cutoff percentage of the bulk composition).

- Second, it is challenging, if not impossible, to replicate the organic content to high fidelity, but we can replicate it to acceptable fidelity for many purposes by using kerogen or sub-bituminous coal instead.

- Third, we do not have adequate data on particle sizing of asteroid regolith **[SKG 6]**, so regolith simulants today must be based upon rough estimations of particle size. Furthermore we have no data on subsurface particle sizing and possible layering of different particle sizes, so we cannot create simulants that can replicate that with any fidelity.

- Fourth, we can use clays that will absorb and release volatiles at a range of thermal and vacuum conditions and we can roughly match the volatile release patterns of meteorite samples, so we believe we can create that with sufficient fidelity to support engineering tests of volatile extraction.

- Fifth, we can control mineralogy to produce the correct magnetic susceptibility of asteroid materials (again based on measurements with meteorites).

Amara Graps | graps@psi.edu

- Sixth, it appears that by mixing the correct minerals, the simulant naturally produces the desired albedo, but we have not yet measured the reflection spectra to discover if the mixing of minerals naturally results in a spectrum that matches space materials.

- Seventh, the judicious choice of the phyllosilicate minerals (among the other correctly chosen minerals), with a carefully designed wetting and re-drying process, has been found adequate to create cobbles that have the desired strength properties, matching the unconfined compressive strength of actual meteorites.

There are many additional properties of asteroids that have not been replicated, but it seems that these should be adequate for most mining tests at least in the early phases of asteroid mining, as long as the users recall the limitations of simulants and do not try to use them for any purpose where they do not match the properties of space materials.

**r. Processing of mined materials will depend on composition and structure of the asteroid, and is a matter of engineering; is it necessary to develop these methods in the near future or can it be postponed until the asteroid mining industry is more mature? was Q24. [revisit]**

This question is behind the long discussion in the "**Mission Design**" discussion. See below.

Another answer. From ASIME-2016:  Yes, in the near future. There is already work on 'processing' if you consider the studies on interactions with the regolith surface: 1) Real, as in space missions and 2) microgravity experiments, and 3) by simulation in granular experiments.

Why structure is important: You need to interact with the object to know how that object will behave and respond mechanically. Composition won't tell you how your sampling tool will interact with the surface. It is more complicated for the reconnaissance missions.  These presentations discussed this topic in great detail:  (Biele et al., 2016; Murdoch, 2016; Michel et al., 2016).

**s. Is anyone working on software that combines various meteorite spectra in an attempt to reproduce an asteroid spectrum that might contain contributions from two or more surface compositions? was Q19.**

P. Michel said that he thinks that there are a lot of people working on this.

The section 3 " Spectral Analysis" in Takir et al., (2019) may be useful.  They keep to the carbonaceous chondrite classes, but explore all sub-types.

Takir et al. (2019)'s work focused on the 2-4 μm spectral region that exhibits diagnostic absorption features, including the water and organics features, which was the C-type asteroids and the carbonaeous chondrite classes.  They measured meteorite reflectance spectra under asteroid-like conditions in the laboratory (asteroid environment temperature and pressure), adding more parameters to the asteroid measurement conditions that other researchers found are critical (Beck: observation geometry, Donaldson-Hanna, thermal gradient).

The 3-μm band was isolated and then divided by a straight-line continuum, which was determined by two reflectance maxima just shortward of the absorption 2.60-2.65 μm and just longward of the absorption 3.5-3.85 μm. To provide a reference wavelength for the asteroid matching, the band depths for the chondrites' spectra were computed at ~2.90 μm relative to the linear continuum that was calculated from a linear regression of data from 1.95-2.5 μm.

They found a good correlation between the 3-μm band area (BA) and 2.90-μm band depth (BD 2.90). Generally, they found a good and qualitative agreement with the petrological classification and the degree of aqueous alteration of carbonaceous chondrites and their 3-μm band parameters (e.g., band area, band center).

> **t. What signatures of past water of hydrated minerals could be observed on an asteroid surface that might indicate subsurface water or hydrated minerals? was Q16. [revisit]**

The presentations: Barucci (2018-ASIME), Rivkin (2018-ASIME), and Campins (2018-ASIME) go into many details about water-rich asteroids and NEOs and their detection.

> **u. How can the surface desiccation of carbonaceous asteroids be determined (via remote observation, in situ measurements, or theoretical models) as a function of MBA to NEO transport lifecycle? was Q17. [revisit]**

The presentations: Barucci (2018-ASIME), Rivkin (2018-ASIME), and Campins (2018-ASIME) go into many details about water-rich asteroids and NEOs and their detection.

Theoretical models for computing ice recession on main-belt asteroids.
Modeling the past orbital history of each NEO would show which ones went close to the Sun for long enough to bake out any free water. [**SKG 5**]  Lab experiments have shown that hydroxyl in clays can survive vacuum conditions and require fairly high temperatures to be released quickly from the clay, on the order of 400 to 500 deg C. More work is needed to determine if it can be released slowly at lower temperatures in vacuum. Current best estimate is that unless a carbonaceous asteroid was close enough to the sun to reach fairly high temperatures at or beneath the surface, then the clays at or beneath the surface should still be hydrated. This agrees with the observation of meteorites that are still hydrated when reaching Earth's surface despite being this near to the sun.

Amara Graps | graps@psi.edu

There are theoretical models for computing ice recession on main-belt asteroids (e.g., Schorghofer, 2008, 2016; Prialnik & Rosenberg, 2009). In principle, these could be combined with dynamical modeling of MBA-NEO pathways to determine the integrated total ice loss for an object over its entire migration to near-Earth space. The remaining ice content will depend on things like the "starting" ice content of the originating main-belt asteroid and the specific thermal properties of the object (which can be estimated, but may or may not be precisely correct for any particular object) though. This will also necessarily be a statistical assessment given the chaotic nature of the dynamical transition of a MBA to a NEO.

# References

**A. Barucci** (**2018-ASIME**) Overview of the Asteroid composition : water and mineralogy theme. Spitzer Rosetta Lutetia flyby results with spectral limitations.
Video: https://youtu.be/7GQgJscCdFE?t=5h25m16s
Presentation: http://geophy.uni.lu/users/tonie.vandam/asime-2018/presentations/barucci.pdf.gz

**P. Beck** and L. Bonal (**2018-ASIME**). Quantifying hydration from IR signatures of primitive meteorites.
Presentation:http://geophy.uni.lu/users/tonie.vandam/asime-2018/presentations/beck.pdf.gz

**J. Biele**, Ulamec, Stephan; Maurel, Clara; Michel, Patrick; Ballousz, Ronald;, Grundmann, Jan Thimo (2016). Touching an asteroid - mechanics of interaction. Presentation at the Asteroid Science Intersections with Mine Engineering Conference, Luxembourg. September 21-22, **2016**.

**R. Binzel** (**2019**). "Small bodies looming large in planetary science", Nature Astronomy vol. 3, April 2019, pp. 282–283.

**D. T. Blewett** & W. Denevi, Brett & Le Corre, Lucille & Reddy, Vishnu & E. Schröder, Stefan & M. Pieters, Carle & Tosi, Federico & Zambon, Francesca & De Sanctis, Maria Cristina & Ammannito, E & Roatsch, Thomas & A. Raymond, Carol & T. Russell, Christopher. (**2015**). Optical Space Weathering on Vesta: Radiative-transfer Models and Dawn Observations. Icarus. 265. 10.1016/j.icarus.2015.10.012.

**L, Bona**l and P. Beck (**2018-ASIME**). Heating processes in primitive asteroids as revealed by the study of organics and hydration of CMs and ungrouped C1/2 chondrites.
Presentation:http://geophy.uni.lu/users/tonie.vandam/asime-2018/presentations/bonal.pdf.gz

**W. F. Bottke,** D. Vokrouhlický, K. J. Walsh, M. Delbo, P. Michel, D. S. Lauretta, H. Campins, H. C. Connolly Jr., D. J. Scheeres, S. R. Chelsey. (**2015**). "In search of the source of asteroid (101955) Bennu: Applications of the stochastic YORP model." Icarus 247, 191–217. doi:10.1016/j.icarus.2014.09.046

**W. F. Bottke Jr.,** D. Durda, D. Nesvorny, R. Jedicke, A. Morbidelli, D. Vokrouhlicky, H. Levison. (**2005**). The fossilized size distribution of the main asteroid belt. Icarus 175, 111– 140 (2005).
doi:10.1016/j.icarus.2004.10.026

Amara Graps | graps@psi.edu

**W.F. Bottke,** A. Morbidelli, R. Jedicke, J.M. Petit, H.F. Levison, P. Michel, and T.S. Metcalfe (**2002**). "Debiased orbital and absolute magnitude distribution of the near-Earth objects". Icarus, 156(2), pp.399-433.

**D. Britt, P. Metzger (\*), et al (2016)** Requirements for Families of Asteroid Regolith Simulants. Presentation at the Asteroid Science Intersections with Mine Engineering Conference, Luxembourg. September 21-22, 2016.

**R. Brunetto** and C. Lantz (**2019**). "Laboratory perspectives on sample returns from Hayabusa2 and OSIRIS-REx", Nature Astronomy, vol 3, Apr 2019, pp. 290–292.

**R. Brunetto,** Loeffler, M.J., Nesvorný, D., Sasaki, S. and Strazzulla, G., (**2015**). Asteroid surface alteration by space weathering processes. Asteroids IV, pp.597-616.

**C.R. Chapman, (2004).** Space Weathering of Asteroid Surfaces. Annu. Rev. Earth Planet. Sci., 32, pp.539-567.

**B.E. Clark**, Hapke, B., Pieters, C. and Britt, D., (**2002**). Asteroid space weathering and regolith evolution. Asteroids III, 585.

**B. Carry** Asteroid Science Intersections with In-Space Mine Engineering (ASIME) (**2018-ASIME**). "The composition of asteroids from sky surveys". Video: https://youtu.be/9SX4nFnIN4M?t=1h11m47s
Presentation: http://geophy.uni-lu/users/tonie.vandam/asime-2018/presentations/carry.pdf.gz

**H. Campins** (**2018-ASIME**). Spectral Diversity Among Primitive Asteroids: Inner Belt Families. Video: https://youtu.be/7GQgJscCdFE?t=5h54m18s and
Presentation:http://geophy.uni.lu/users/tonie.vandam/asime-2018/presentations/campins.pdf.gz
See also: **H. Campins**, de León, J., Licandro, J., et al., **(2018).** Chapter 5: Compositional diversity among primitive asteroids. Book: Primitive Meteorites and Asteroids: Physical, Chemical and Spectroscopic Observations Paving the Way for Exploration. Elsevier. p345-367 https://doi.org/10.1016/B978-0-12-813325-5.00005-7

**M. Delbo**, (**2016**). Remote sensing investigation of asteroids: the nature of the regolith revealed by the thermal inertia. ASIME 2016: Asteroid Intersections with Mine Engineering, Luxembourg. September 21-22, 2016.

**M. Delbo**, Libourel, G., Wilkerson, J., Murdoch, N., Michel, P., Ramesh, K.T., Ganino, C., Verati, C. and Marchi, S., (**2014**). Thermal fatigue as the origin of regolith on small asteroids. *Nature* Vol 508, pages 233–236.

**J. de Leon, (2018-ASIME).** Ground-based spectroscopy in the visible and infrared to extract mineralogical composition of asteroids**.** Video: https://youtu.be/7GQgJscCdFE?t=2h14m47s  Presentation: http://geophy.uni-lu/users/tonie.vandam/asime-2018/presentations/deleon.pdf.gz

**J. de León,** H. Campins, D. Morate, M. De Prá, V. Alí-Lagoa, J. Licandro, J.L. Rizos, N. Pinilla-Alonso, D.N. DellaGiustina, D.S. Lauretta, M. Popescu, V. Lorenzi **(2018)** "Expected spectral characteristics of (101955) Bennu and (162173) Ryugu, targets of the OSIRIS-REx and Hayabusa2 missions", Icarus 313. p. 25–37.

Amara Graps | graps@psi.edu

**F.E. DeMeo, and B. Carry** (**2014**). Solar System evolution from compositional mapping of the asteroid belt. Nature, 505(7485), pp.629-634.

**F.E. DeMeo and B. Carry** (**2013**). "The taxonomic distribution of asteroids from multi-filter all-sky photometric surveys", Icarus Volume 226, Issue 1, September–October 2013, Pages 723-741. https://doi.org/10.1016/j.icarus.2013.06.027

**F.E. DeMeo**, R.P. Binzel, S.M. Slivan, S.J. Bus (2009). An extension of the Bus asteroid taxonomy into the near infrared. Icarus 202 (July), 160–180. https://doi.org/10.1016/j.icarus.2009.02.005

**W. Denevi**, Brett & S. Robinson, Mark & Boyd, A & Sato, Hiroyuki & W. Hapke, Bruce & Ray Hawke, B. (**2014**). Characterization of space weathering from Lunar Reconnaissance Orbiter Camera ultraviolet observations of the Moon. Journal of Geophysical Research: Planets. 119. https://doi.org/10.1002/2013JE004527 .

**A.J. Dombard** and Freed, A.M., (**2002**). Thermally induced lineations on the asteroid Eros: Evidence of orbit transfer. Geophysical research letters,29(16).

**K.L. Donaldson-Hanna** , N. E. Bowles, H. C. Connolly Jr., V. E. Hamilton, L. P. Keller, D. S. Lauretta, L. F. Lim, and D. S. Schrader, (**2018**). "Analogue Materials Measured Under Simulated Asteroid Conditions" **2018-ASIME**. Video: https://youtu.be/9SX4nFnIN4M (*only last ½*) Presentation:http://geophy.uni.lu/users/tonie.vandam/asime-2018/presentations/donaldson.pdf.gz

**L. Drube** and A. W. Harris (**2016**). A new method of identifying metal-rich asteroids. ASIME 2016: Asteroid Intersections with Mine Engineering, Luxembourg. September 21-22, 2016

**R. Duffard.** Determination of basaltic and hydrated material from remote sensing techniques. ASIME 2016: Asteroid Intersections with Mine Engineering, Luxembourg. September 21-22, **2016**

**M. Elvis, A. Taylor, A. Stark and P. Vereš (2018).** Real Life Testing of a Novel Astronomical Prospecting Technique. Video: https://youtu.be/7GQgJscCdFE?t=2h44m39s Presentation:http://geophy.uni.lu/users/tonie.vandam/asime-2018/presentations/elvis.pdf.gz

**M. Elvis** (**2016**). What can space resources do for astronomy and planetary science?. Space Policy, 37, pp.65-76.https://arxiv.org/abs/1608.01004

**A. Fitzsimmons** Asteroid Science Intersections with In-Space Mine Engineering (**2018-ASIME**). "First Day Wrap-up". April 16 Video: https://youtu.be/7GQgJscCdFE?t=8h46m28s Presentation: http://geophy.uni.lu/users/tonie.vandam/asime-2018/presentations/Fftzsimmons.pdf.gz

**J.L. Galache** (**2016**). Asteroid Mining: State of the (Scientific) Art. ASIME 2016: Asteroid Intersections with Mine Engineering, Luxembourg. September 21-22, 2016.

Amara Graps | graps@psi.edu

**J.L Galache;** Beeson, C.L.; McLeod, K.K.; Elvis, M. (**2015**). The need for speed in Near-Earth Asteroid characterization, Planetary and Space Science, Volume 111, June 2015, Pages 155-166, ISSN 0032-0633, http://dx.doi.org/10.1016/j.pss.2015.04.004 .

**M. Granvik**, M., Morbidelli, A., Jedicke, R., Bolin, B., Bottke, W.F., Beshore, E., Vokrouhlický, D., Delbò, M. and Michel, P., (**2016**). Super-catastrophic disruption of asteroids at small perihelion distances. Nature, 530(7590), pp.303-306.

**A. Graps** +30 co-authors (**2016**). ASIME 2016 : "Answers to Questions from the Asteroid Miners". White Paper from the Asteroid Science Intersections with In-Space Mine Engineering (ASIME) 2016 conference on September 21-22, 2016 in Luxembourg City. https://arxiv.org/abs/1612.00709

**A. Graps** (**2018**) Informal collection of notes of the patched charge model in the context of surfaces of porous, airless small planetary bodies: Kempf/Horanyi EPSC 2017 presentation (video and slides), research articles of discovery, and list of potential dust charging mysteries over the decades the patched charge model might address. https://isn.page.link/VY8A

**S.F. Green**. Asteroid Science Intersections with In-Space Mine Engineering (**2018-ASIME**). April 17 "Wrap-Up". Video: https://youtu.be/9SX4nFnIN4M?t=6h59m29s Presentation: http://geophy.uni.lu/users/tonie.vandam/asime-2018/presentations/green.pdf.gz

**S.F. Green** (**2016**). Remote Sensing: Constraining Asteroid Target and Landing Site Selection. ASIME 2016: Asteroid Intersections with Mine Engineering, Luxembourg. September 21-22, 2016.

**S. Greenstreet**, Ngo, H. and Gladman, B., (**2012**). The orbital distribution of near-Earth objects inside Earth's orbit. Icarus, 217(1), pp.355-366.

**T. Grundmann et al (30 co-authors):** DLR Efficient Massively Parallel Prospection for ISRU by Multiple Near-Earth Asteroid Rendezvous using Near-Term Solar Sails and 'Now-Term' Small Spacecraft Solutions, **2018-ASIME**. Video: https://youtu.be/9SX4nFnIN4M?t=6h12m Presentation:http://geophy.uni.lu/users/tonie.vandam/asime-2018/presentations/grundmann.pdf.gz

**B. Gundlach** and J. Blum (**2013**). A new method to determine the grain size of planetary regolith. Icarus, 223(1), pp.479-492.

**Hapke, B., (2001)**. Space weathering from Mercury to the asteroid belt. Journal of Geophysical Research: Planets, 106(E5), pp.10039-10073.

**A.R. Hendrix** and F. Vilas **(2006)**. The Effects of Space Weathering at UV Wavelengths: S-class Asteroids, Astron. J.: 132: 1396-1404.

**H.M Kaluna**, Masiero, J.R. and Meech, K.J., (**2016**). Space weathering trends among carbonaceous asteroids. Icarus, 264, pp.62-71.

Amara Graps | graps@psi.edu

**K.R. Kuhlman**, Sridharan, K. and Kvit, A., (**2015**). Simulation of solar wind space weathering in orthopyroxene. Planetary and Space Science, 115, pp.110-114.

**D. Lauretta**, et al. (**2019**). "The unexpected surface of asteroid (101955) Bennu", Nature https://doi.org/10.1038/s41586-019-1033-6

**P.T. Metzger,** Daniel T. Britt, Stephen D. Covey, and John S. Lewis, "Results of the 2015 Workshop on Asteroid Simulants," Proceedings of Earth and Space 2016: Engineering, Science, Construction, and Operations in Challenging Environments, Orlando, FL, April 11-15, **2016**.

**P. Michel (2016).** Why the Asteroid Impact Mission (AIM) is a crucial step towards asteroid mining? ASIME 2016: Asteroid Intersections with Mine Engineering, Luxembourg. September 21-22, 2016.

**H.-K. Moon** et al. (**2018-ASIME**) A Novel Asteroid Taxonomy with 3D Photometric Colors based on Spectroscopy. Video: https://youtu.be/9SX4nFnIN4M?t=52m56s Presentation: http://geophy.uni-lu/users/tonie.vandam/asime-2018/presentations/moon.pdf.gz

**D. Morate**, J. de León, M. De Prá, J. Licandro, A. Cabrera-Lavers, H. Campins, N. Pinilla-Alonso, V. Alí-Lagoa. (**2016**). Compositional study of asteroids in the Erigone collisional family using visible spectroscopy at the 10.4 m GTC. Astron. Astrophys. 586, A129. doi:10.1051/0004-6361/201527453

**N. Murdoch**, (**2016**). Physical properties of the surface and sub-surface of asteroids. ASIME 2016: Asteroid Intersections with Mine Engineering, Luxembourg. September 21-22, 2016.

**M. Nayak** & Asphaug, E. **(2016).** Sesquinary catenae on the Martian satellite Phobos from reaccretion of escaping ejecta. Nature communications. 7. 12591. 10.1038/ncomms12591.

**C.M. Pieters**, & Noble, S. K. (**2016**). Space Weathering on Airless Bodies. *Journal of geophysical research. Planets*, *121*(10), 1865–1884. doi:10.1002/2016JE005128

**M. Popescu**, J. Licandro, D. Morate, J. de León, D. A. Nedelcu, R. Rebolo, R. G. McMahon, E. Gonzalez-Solares and M. Irwin (**2016**). Near-infrared colors of minor planets recovered from VISTA-VHS survey (MOVIS), A&A, 591, A115. DOI: https://doi.org/10.1051/0004-6361/201628163

**D. Prialnik** and Rosenberg, E.D., (**2009**). Can ice survive in main-belt comets? Long-term evolution models of comet 133P/Elst-Pizarro. Monthly Notices of the Royal Astronomical Society: Letters, 399(1), pp.L79-L83.

**A. Rivkin*** and F. E. DeMeo (**2018-ASIME**). Asteroid composition: mineralogy and water: Part III. How Many Hydrated NEOs Do We Expect? Video: https://youtu.be/7GQgJscCdFE?t=6h15m48s Presentation:http://geophy.uni-lu/users/tonie.vandam/asime-2018/presentations/rivkin.pdf.gz

**A. Rivkin** (**2016**) Water and Hydroxyl in Asteroids: What Do We See and Where Do We See It? ASIME-2016: Asteroid Intersections with Mine Engineering, Luxembourg. September 21-22, 2016.

**S. Sugita, et al. (2019).** "The geomorphology, color, and thermal properties of Ryugu: Implications for parent-body processes." Science https://doi.org/10.1126/science.aaw0422

**D. Takir; K. R. Stockstill-Cahill; C. A. Hibbitts, Y. Nakauchi** (**2019**). 3-μm Reflectance Spectroscopy of Carbonaceous Chondrites under Asteroid-like Conditions. arXiv preprint https://arxiv.org/pdf/1904.09453.

**D. Takir, et al. (2013).** Nature and degree of aqueous alteration in CM and CI carbonaceous chondrites. Meteorit. Planet. Sci. 48, 1618–1637**.**

**P. Tricarico**, (**2016**). The near-Earth asteroids population from two decades of observations. arXiv preprint arXiv:1604.06328.

**B.P. Weiss**, Elkins-Tanton, L.T., Antonietta Barucci, M., Sierks, H., Snodgrass, C., Vincent, J.-B., Marchi, S., Weissman, P.R., Pätzold, M., Richter, I., Fulchignoni, M., Binzel, R.P., Schulz, R. (**2012**). Possible evidence for partial differentiation of asteroid Lutetia from Rosetta, Planetary and Space Science Volume 66, Issue 1, June 2012, Pages 137-146

**M. Willman**, Jedicke, R., Moskovitz, N., Nesvorný, D., Vokrouhlický, D. and Mothé-Diniz, T. (**2010**). Using the youngest asteroid clusters to constrain the space weathering and gardening rate on S-complex asteroids. Icarus,208(2), pp.758-772.

**Wu Yunzhao** (**2018-ASIME**) In-situ spectra from Chang'E-3 and laboratory spectra of meteorites
Video: https://youtu.be/9SX4nFnIN4M?t=15m16s
Presentation: http://geophy.uni.lu/users/tonie.vandam/asime-2018/presentations/wu.pdf.gz

Amara Graps | graps@psi.edu

# Mission Design Discussion
===========================
Key 25 minute Discussion on Green's Mission Flowchart
ASIME 2018, April 17, 2018, 17:30 CET.
University of Luxembourg, Belval.
===========================

Paraphrasing words, Amara Graps, August 2018.

Video recording for that specific discussion : at 8h21m

Graphic for discussion is below:

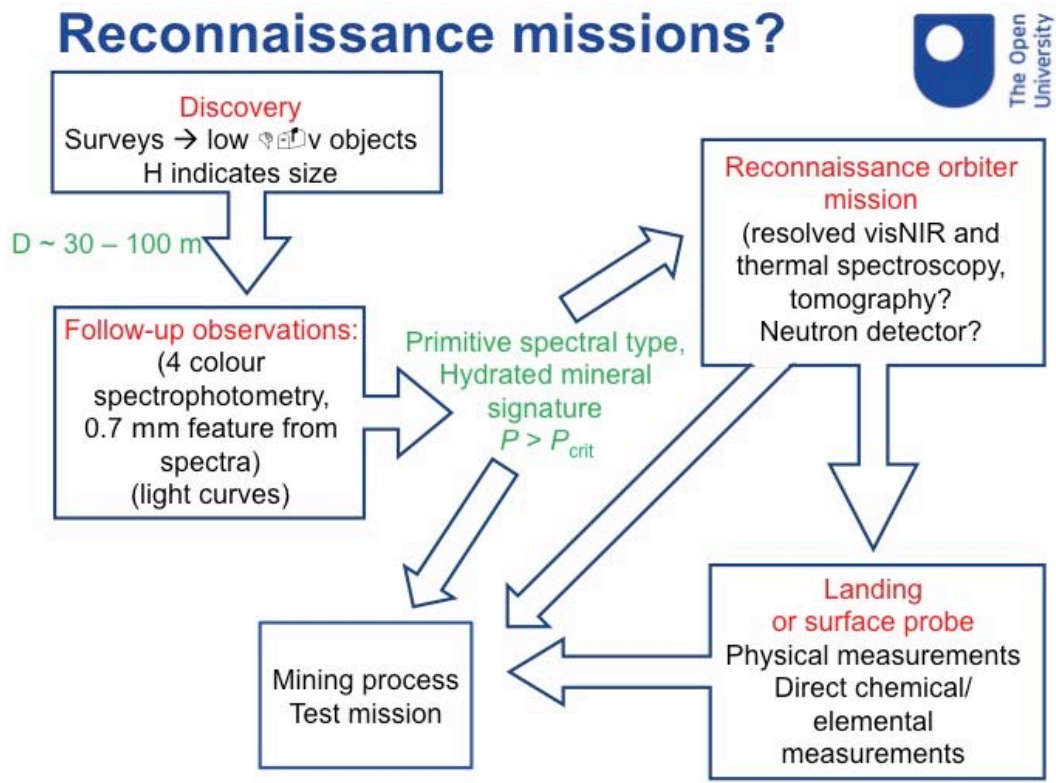

**Green:**
Scientists think of missions as stages
Industry people don't necessarily design missions that way.
They think "what is the quickest way to make a profit?"

1) you want to have some level of confidence about the target of your mission that it will lead you to a profit.

2) Discover the object

Amara Graps | graps@psi.edu

Right size?

Follow up observations to a viable target

If Primitive:

You could go there and study and....your mission is very expensive.

OR

You could send your mining mission directly there. Or send three to three different objects.

You might choose to make a reconnaissance orbit mission and then land on.

But you might still not get the answers you want.

The Mining mission may still be cheaper.

The DISCOVERY is the most important.

The arrows point to engineering questions, not science questions.

**Abell:** This is great. The number one, foundational here is the Discovery. Which needs a survey for low-delta-v objects.

**Rivkin**: So to echo what you said: which of those arrows, going from the top to the middle or wherever, is an engineering question, not a science question. Who are the engineers who could provide additional insight into this.

**Michel**:
30m size is borderline.. asteroid could be a really fast spinner

**Abbud**: As an engineer, I like the graph.

**Alan Fitzsimmons** : Discovery with NEO Cam could give you all data and not need follow-on Observations.

**Raymond**: Scientists can continue identifying targets.

Mining community could provide some science data to the scientists.

The two could work more synergestically.

**Michel**: I agree.

The knowledge gained by interaction with the surface is precious.

**Elvis** : be careful- that is the property of the miners.

We need more engineers/miners at the next ASIME.

**Bonin** (DSI's perspective) :

An intermediate step in our roadmap is predicated on sample return. As low cost sample return that we can achieve. Which means that the samples are not going to be truly pristine. We would go the Mining process Test mission box as quickly as possible.

The reason that we want to do that is that having a priori immaculate knowledge is going to take too long. We prioritize doing a large number of missions because we expect a high failure rate. We recognize it's a long game.

I'm interested in the feedback from people in the room for this kind of approach. That's how we think about it.

We want to go directly to Test missions with interacting with the surface right away with multiple missions, expecting a high failure rate.

**Green's** response: A dirty asteroid sample is better than no sample. The sample inside may be good.

**Michel**: IP: Most of the knowledge by the Miners is from taxpayer money.  Plus science is universal.

**Bonin**: That knowledge is so commercial and valuable.. we want to monetize that knowledge.

**Sercel**: I recently spent time at a goldmine in Peru. They would not give anyone in the public that data.

However, if DSI has that asteroid data, I would advocate that they give it to the public (joking).

Also that flowchart is a traditional flowchart.

**Grundmann**: We are still at the Lewis and Clark stage of surveying. Surveying benefits all companies equally. I would advocate the public focus on the two boxes on the right in a kind of public-private partnership. The companies can jump directly to the Mining missions.

**Green**: I see these boxes representing mission much larger than what  you're doing with your Gossamer solar sail. The main issue is time.

In business, time is money.

**Sercel**: The appropriate paradigm for asteroid mining is FISHING, not prospecting. Therefore you go from Discovery to follow up observations, and then directly to a full-scale mining mission.

**Galache**: (half-joking to Green) I'm 'disappointed' in your charts- you are killing the business case for my company, which is about reconnaissance orbiter missions.

If we are going to do reconnaissance missions, then we are going to have to have ground truth. Once we do a few of them, then that will allow us to revisit all of the other spectra and be able to reinterpret the data.

And I'm sorry, Patrick, but that is very valuable data, not to freely give away.

**K. Moon**: Back to issue of Discovery:

The Aten group of asteroids should be the lowest delta-v objects. But it is very difficult to discover and to follow/characterize. I suggest to use a network of telescopes, which can follow up. At the same time to use Space Telescope such as Gaia.

**Rivkin**: We are the dawn of a new large economy and we are here to help guide it and see how and where it goes. We would like to help the maximum number of people.

If we are all in this large community together, then we can make sure that everyone gets what they need. Certainly some data should be proprietary for the companies or else the economical benefits won't work and the market is the market.

The scientists also have something valuable that they are offering. If the scientists are doing something for free, eventually they will get to a point that they are wondering why are they doing that if nothing is coming back to them.

So I think we can all work this together.

**Bonin**: I think our objectives are all complementary, fundamentally.

I.e. Your objectives 1, 2, and 3, are probably my objectives 6, 7, and 8. Our commercial interests and your scientific interests are complementary.

**Sercel**: I would agree with that also. I think that there are questions that the scientists have that the government is willing to pay for that will have peripheral benefit to the companies. And that there are questions that the miners have that they need the scientists' help for, which they will pay for.

Most geologists are not employed in academia today. They are employed in prospecting for fossil fuels and minerals. I think mining is a wonderful job growth opportunity for the scientific community.

On this chart: a couple of edits:

1) we don't see that H indicates size, until you have done the observation and see what the albedo is.

**Green**: You use H as a guide so that you don't go after something that isn't what you want. It's for narrowing down the size range. Of course you get the true size when you measure the albedo.

Amara Graps | graps@psi.edu

**Sercel**: Edit 2: I would put the low end of the diameter scale much lower than 30m. There might be 100 tons of ice in a 10-meter object.

**Green**: Right. But it depends if you are prepared to go to that 10-meter object with no follow up. And you might find when you get there that it is spinning at 10 revolutions per second. You can't follow it up. It is too fast.

**Sercel**: Yes, follow up observations need to be done in such a way that it gives you such information as spin. I recommend that you read our Sutter report and see what mistakes we made. We do need to know about spin as part of the follow up observation.

**Green**: You can't fight photons. I worried that 30 meter was too small. That size is difficult to observe. Ten meter is definitely too small. That is, too small to get access to the big telescopes with enough observing time to determine physical characteristics.

**Sercel**: Our Sutter report (not on microphone) should be accessible.

# Program with links to recordings and presentations

## Day One:  16 April 2018  8:00-22:00

- **8:00 Registration opens. Coffee/Tea/Refreshments**
- **8:30 ASIME 2018 Introductory Remarks**
  - *Tonie Van Dam, University of Luxembourg*  Video: https://youtu.be/7GQgJscCdFE?t=8m54s
- **8:40-9:30  I. Questions from the Asteroid Mining companies**
  - *Patrick Michel + Amara Graps* present the Asteroid Scientific Composition **Questions** from the companies: *Deep Space Industries, Planetary Resources, TransAstra, Aten Engineering, SolSys Mining, ispace*
    - Video: https://youtu.be/7GQgJscCdFE?t=9m29s
    - Presentation:http://geophy.uni.lu/users/tonie.vandam/asime-2018/presentations/graps-michel.pdf.gz

### 9:30-9:40 BREATHING BREAK (DIGNITARIES + PRESS ARRIVE)

- **9:40 Opening Remarks**
  - Opening: *Etienne Schneider*, Deputy Prime Minister and Minister of the Economy, The Government of Grand Duchy of Luxembourg Video: https://youtu.be/7GQgJscCdFE?t=1h12m20s
  - University of Luxembourg Overview:  **Yves Elsen**, *Chairman of the Board of Governors of the University of Luxembourg* Video: https://youtu.be/7GQgJscCdFE?t=1h24m29s
  - Luxembourg National Research Fund (FNR) Overview:  **Marc Schiltz**, *Executive Head of the FNR* Video: https://youtu.be/7GQgJscCdFE?t=1h31m54s
  - ASIME 2018 Scientific Overview:  **Patrick Michel**, *Senior Researcher, Observatoire de la Côte d'Azur, CNRS, FR* Video: https://youtu.be/7GQgJscCdFE?t=1h42m13s

### 10:00-10:20 COFFEE BREAK

- **10:20 II. Composition from Spectroscopic Observations from the Ground**
  - 10:20-11:20 Keynote: ***Julia de Leon et al.***, *Instituto de Astrofísica de Canarias - IAC.*  Ground-based spectroscopy in the visible and near infrared to extract mineralogical composition of asteroids
    - Video: https://youtu.be/7GQgJscCdFE?t=2h14m47s
    - Presentation:http://geophy.uni.lu/users/tonie.vandam/asime-2018/presentations/deleon.pdf.gz
  - 11:20-11:40 **Martin Elvis, Anthony Taylor, Anthony Stark and Peter Vereš** *Harvard-Smithsonian Center for Astrophysics* Real Life Testing of a Novel Astronomical Prospecting Technique
    - Video: https://youtu.be/7GQgJscCdFE?t=2h44m39s
    - Presentation:http://geophy.uni.lu/users/tonie.vandam/asime-2018/presentations/elvis.pdf.gz
  - 11:40-12:00 ***Dan Britt and Kevin Cannon**** *University of Central Florida* Simulating asteroid materials with realistic compositions
    - Video: https://youtu.be/7GQgJscCdFE?t=3h3m50s
    - Presentation:http://geophy.uni.lu/users/tonie.vandam/asime-2018/presentations/cannon.pdf.gz
- **12:00-12:45 Intro/ Status / Pitches for company asteroid mining activities: *Planetary Resources, Deep Space Industries, TransAstra, Aten Engineering,  SolSys Mining***
    - Videos: https://youtu.be/7GQgJscCdFE?t=3h36m39s  Note: DSI was not recorded.
    - Presentations:
      - Planetary Resources: http://geophy.uni.lu/users/tonie.vandam/asime-

2018/presentations/Planetary_Resources.pdf.gz
- Trans Astra: http://geophy.uni.lu/users/tonie.vandam/asime-2018/presentations/TransAstra.pdf.gz
- Aten Engineering
  http://geophy.uni.lu/users/tonie.vandam/asime-2018/presentations/Aten_Engineering.pdf.gz
- Sol Sys Mining
  http://geophy.uni.lu/users/tonie.vandam/asime-2018/presentations/SolSys_Mining.pdf.gz

## 12:45-13:45 LUNCH

- **13:45 II. continued Composition from Spectroscopic Observations from the Ground**
  - 13:45-14:15 Keynote: *Antonella Barucci* *Observatoire Paris-Site de Meudon* Overview of the Asteroid composition : water and mineralogy theme.  Spitzer Rosetta Lutetia flyby results with spectral limitations.
    - Video: https://youtu.be/7GQgJscCdFE?t=5h25m16s
    - Presentation: http://geophy.uni.lu/users/tonie.vandam/asime-2018/presentations/barucci.pdf.gz
  - 14:15-14:45 Keynote: **Humberto Campins** *University of Central Florida* Asteroid composition: mineralogy and water: Part II. Compositional Diversity Among Primitive Asteroids
    - Video: https://youtu.be/7GQgJscCdFE?t=5h54m18s
    - Presentation:http://geophy.uni.lu/users/tonie.vandam/asime-2018/presentations/campins.pdf.gz
  - 14:45-15:15 Keynote: *Andy Rivkin\** *John Hopkins University* **and F. E. DeMeo** *MIT* Asteroid composition: mineralogy and water: Part III. How Many Hydrated NEOs Do We Expect?
    - Video: https://youtu.be/7GQgJscCdFE?t=6h15m48s
    - Presentation:http://geophy.uni.lu/users/tonie.vandam/asime-2018/presentations/rivkin.pdf.gz

## 15:15-15:35 COFFEE BREAK

- 15:35-15:50 *Angel Abbud-Madrid and Christopher Dreyer* *Colorado School of Mines* Technology and Asteroid Science Working Together for the Successful Development of Asteroid Resources
  - Video: https://youtu.be/7GQgJscCdFE?t=7h10m29s
  - Presentation: http://geophy.uni.lu/users/tonie.vandam/asime-2018/presentations/abbud-madrid.pdf.gz
- 15:50-16:10 *Sampsa Pursiainen and Mika Takala* *Tampere University of Technology, FI* Computational considerations for 3D full-wave asteroid tomography
  - Video: https://youtu.be/7GQgJscCdFE?t=7h29m56s
  - Presentation: http://geophy.uni.lu/users/tonie.vandam/asime-2018/presentations/pursiainen_takala.pdf.gz
- **16:10  III. Lunar and other Space Resources**
  - 16:10-16:40 Keynote: *Ian Crawford* The Moon's Role in the Development of Space Resources
    - Video: https://youtu.be/7GQgJscCdFE?t=7h53m1s
    - Presentation: http://geophy.uni.lu/users/tonie.vandam/asime-2018/presentations/crawford.pdf.gz
  - 16:40-16:50 *Tom Wirtz* *Luxembourg Institute of Science and Technology (LIST)* LIST Applications and Activities for Space Resource Utilisation
    - Video: https://youtu.be/7GQgJscCdFE?t=8h20m15s
    - Presentation:http://geophy.uni.lu/users/tonie.vandam/asime-2018/presentations/wirtz.pdf.gz

together with:

- o 16:50-17:00 **Abigail Calzada-Diaz, Kyle Acierno, and Philippe Ludivig:** *ispace* <u>The ispace's Approach to Lunar Resources Exploration</u>
  - ▪ Video: https://youtu.be/7GQgJscCdFE?t=8h31m16s
  - ▪ Presentation: http://geophy.uni.lu/users/tonie.vandam/asime-2018/presentations/calzada-diaz.pdf.gz

**17:00-18:00 Round table lead at the end of the first day**. *Alan Fitzsimmons* *Queen's Univ Belfast Astrophysics Research Centre*

- o Video: https://youtu.be/7GQgJscCdFE?t=8h46m28s
- o Presentation:http://geophy.uni.lu/users/tonie.vandam/asime-2018/presentations/Fftzsimmons.pdf.gz

**19:00-22:00 WALKING DINNER**

**19:15-19:30 Dinner speech:** *Pete Worden*, *Chairman, Breakthrough Prize Foundation & Member of the Luxembourg Space Resources Advisory Board* <u>'Search for Life'</u>

# Day Two: 17 April 2018 8:00-21:00

- **8:00 Coffee/Tea/Refreshments**
- **8:30 IV. Asteroid Composition from Lab measurements**
  - o 8:30-9:20 two Keynotes: **Lydie Bonal and Pierre Beck** *University of Grenoble* <u>Heating processes in primitive asteroids as revealed by the study of organics and hydration of CMs and ungrouped C1/2 chondrites</u>) *Next two videos were not recorded.*
    - ▪ Presentation:http://geophy.uni.lu/users/tonie.vandam/asime-2018/presentations/bonal.pdf.gz

  and

  - o **Pierre Beck, Lydie Bonal et al.** *University of Grenoble* <u>Quantifying hydration from IR signatures of primitive meteorites</u>
    - ▪ Presentation:http://geophy.uni.lu/users/tonie.vandam/asime-2018/presentations/beck.pdf.gz
  - o 9:20-9:40 **Kerri L. Donaldson-Hanna et al.** *University of Oxford* <u>Analogue Materials Measured Under Simulated Asteroid Conditions: Insights into the Interpretation of Thermal Infrared Remote Sensing Observations</u>
    - ▪ Video: https://youtu.be/9SX4nFnIN4M (*only last ½*)
    - ▪ Presentation:http://geophy.uni.lu/users/tonie.vandam/asime-2018/presentations/donaldson.pdf.gz
  - o 9:40-10:00 **Wu Yunzhao** *Key Laboratory of Planetary Sciences, Purple Mountain Observatory, Chinese Academy of Sciences* <u>In-situ spectra from Chang'E-3 and laboratory spectra of meteorites</u>
    - ▪ Video: https://youtu.be/9SX4nFnIN4M?t=15m16s
    - ▪ Presentation: http://geophy.uni.lu/users/tonie.vandam/asime-2018/presentations/wu.pdf.gz

**10:00-10:20 COFFEE BREAK**

- **10:20 V. Composition from Taxonomy with Dynamics**
  - o 10:20-10:40 **H.-K. Moon et al.** *Korea Astronomy and Space Science Institute* <u>A Novel Asteroid Taxonomy with 3D Photometric Colors based on Spectroscopy</u>
    - ▪ Video: https://youtu.be/9SX4nFnIN4M?t=52m56s
    - ▪ Presentation:http://geophy.uni.lu/users/tonie.vandam/asime-2018/presentations/moon.pdf.gz
  - o 10:40-11:10 Keynote: **Benoît Carry** *Observatoire de la Côte d'Azur* <u>The composition of</u>

asteroids from sky surveys
- Video: https://youtu.be/9SX4nFnIN4M?t=1h11m47s
- Presentation:http://geophy.uni.lu/users/tonie.vandam/asime-2018/presentations/carry.pdf.gz

- **11:10  VI. Composition from Space Missions**
  - 11:10-11:40 Keynote: *Paul Abell* for *Amy Mainzer*  JPL (topic: NEOWise)
    - Video: https://youtu.be/9SX4nFnIN4M?t=1h40m48s
    - Presentation:http://geophy.uni.lu/users/tonie.vandam/asime-2018/presentations/mainzer.pdf.gz
  - 11:40-12:10 Keynote: *Tomoki Nakamura*       *Tohoku University*  Japanese Second Sample Return Mission: Hayabusa 2 . Note: Nakamura not recorded.

## 12:10-13:40 LUNCH

  - 13:40-14:20 Keynote: *Michael Küppers*, *Ian Carnelli,* *ESA*, *Patrick Miche*l  *Observatoire de la Côte d'Azur* Hera mission relevance for asteroid resource exploitation
    - Video: https://youtu.be/9SX4nFnIN4M?t=4h10m53s
    - Presentation:http://geophy.uni.lu/users/tonie.vandam/asime-2018/presentations/michel.pdf.gz
  - 14:20-14:50 Keynote: *Carol Raymond*  JPL (topic: Dawn)
    - Video: https://youtu.be/9SX4nFnIN4M?t=4h34m22s
    - Presentation:http://geophy.uni.lu/users/tonie.vandam/asime-2018/presentations/raymond.pdf.gz
  - 14:50-15:10 *Lin Yangting* *Key Laboratory of Earth and Planetary Physics, Institute of Geology and Geophysics, China.*  Possible Space Resources and Potential Applications in Future
    - Video: https://youtu.be/9SX4nFnIN4M?t=5h14m40s
    - Presentation: http://geophy.uni.lu/users/tonie.vandam/asime-2018/presentations/lin.pdf.gz

## 15:10-15:30 COFFEE BREAK

  - 15:30-15:50 *Grundmann et al* (30 co-authors): *DLR* Efficient Massively Parallel Prospection for ISRU by Multiple Near-Earth Asteroid Rendezvous using Near-Term Solar Sails and 'Now-Term' Small Spacecraft Solutions
    - Video: https://youtu.be/9SX4nFnIN4M?t=6h12m
    - Presentation:http://geophy.uni.lu/users/tonie.vandam/asime-2018/presentations/grundmann.pdf.gz
  - 15:50-16:10 *M. Agnan, D. Hestroffer, et al.* Extensive exploration of small bodies with autonomous navigation. BIRDY.
    - Video: https://youtu.be/9SX4nFnIN4M?t=6h40m5s
    - Presentation:http://geophy.uni.lu/users/tonie.vandam/asime-2018/presentations/agnan-hestroffer.pdf.gz

## 16:10-18:00 VII. Wrap-up: How to Improve Our Knowledge *Simon Green* *Open University*

  - Video: https://youtu.be/9SX4nFnIN4M?t=6h59m29s
  - Presentation: http://geophy.uni.lu/users/tonie.vandam/asime-2018/presentations/green.pdf.gz
  - Q & A Panel and White Paper/ Journal Article Discussion & Collection of Questions and Answers

## 18:00-21:00 WALKING DINNER

Amara Graps | graps@psi.edu

Financial support for the ASIME 2018 conference was provided by the following entities.

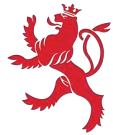

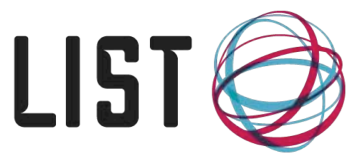

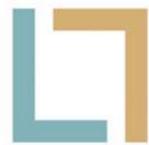

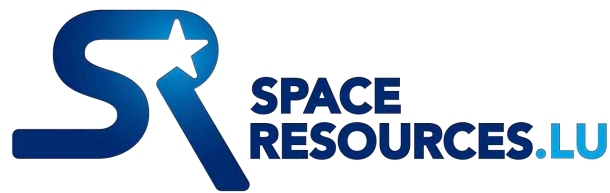

Amara Graps | graps@psi.edu